\documentclass{svjour2}  

\usepackage{bbold} 
\usepackage{amsmath}
\usepackage{graphicx}
\usepackage{array}

\journalname{Few-Body Syst}

\title{Deuteron disintegration in three dimensions\thanks{Dedicated to Prof. H. Wita{\l}a on the occasion of his
  60th birthday}}

\author{K.~Topolnicki \and J.~Golak \and R.~Skibi\'nski \and A.E.~Elmeshneb \and W.~Gl\"{o}ckle \and A.~Nogga \and H.~Kamada}
\institute{
K.~Topolnicki \and J.~Golak \and R.~Skibi\'nski \and  A.E.~Elmeshneb 
\at M. Smoluchowski Institute of Physics, Jagiellonian University, PL-30059 Krak\'ow, Poland 
\and W.~Gl\"{o}ckle
\at Institut f\"ur Theoretische Physik II, Ruhr-Universit\"at Bochum, D-44780 Bochum, Germany
\and A.~Nogga 
\at Forschungszentrum J\"ulich, Institut f\"ur Kernphysik, Institute for Advanced Simulation and J\"ulich Center for Hadron Physics, D-52425 J\"ulich, Germany
\and H.~Kamada
\at Department of Physics, Faculty of Engineering, Kyushu Institute of Technology, 1-1 Sensuicho Tobata, Kitakyushu 804-8550, Japan
}

\def\a{\kern+.6ex\lower.42ex\hbox{$\scriptstyle \iota$}\kern-1.20ex a}
\def\e{\kern+.5ex\lower.42ex\hbox{$\scriptstyle \iota$}\kern-1.10ex e}

\begin{document}

\maketitle

\begin{abstract}

We compare results from the traditional partial wave treatment of deuteron electro-disintegration with a new approach that uses three dimensional formalism. 
The new framework for the two-nucleon (2N) system using a complete set of isospin - spin states made it possible to construct simple implementations that employ a very general operator form of the current operator and 2N states. 

\end{abstract}

\section{Introduction}

The theoretical description of electromagnetic processes is based, like most nuclear physics formalisms, on a partial wave decomposition of relevant operators. This restricts theoretical methods to systems where a relatively small number of partial waves is important. Recently, several formalisms for the three dimensional description of few-body systems and processes have been developed.
In this paper we fill a gap in the development of the three dimensional framework and present an approach that allows for a simple implementation of three dimensional electromagnetic currents. Our final expressions can be translated to a numerical implementation via direct substitutions of $16$ dimensional square matrices representing operators in the 2N isospin - spin space. Finding the matrix representation of relevant operators is greatly simplified by using symbolic programming in the Mathematica $^{\mbox{\textregistered}}$ \cite{math} software package. Our approach allows us to use a very general operator form of current operators and can therefore be used for a wide class of processes. 

In this paper we apply it to the case of electron induced deuteron disintegration and compare the results with traditional partial wave calculations. It is worth noting that the methods presented in the following sections can be applied to the description of other processes, where electroweak probes interact with the 2N system.
Electromagnetic currents can be replaced by any operators acting on the same degrees of freedom; this makes our implementations also useful for calculations involving for example muon capture or neutrino induced deuteron disintegration, performed recently still with the use partial wave expansion \cite{Marcucci:2010ts,Shen:2012xz}.

\section{Formalism and notation}

We adopt a notation in which capital letters describe the total momentum of a two particle system ($\mathbf{P} = \mathbf{p}_{1} + \mathbf{p}_{2}$), lower-case letters describe the relative momentum ($\mathbf{p} = \frac{1}{2} \left(\mathbf{p}_{1} - \mathbf{p}_{2}\right)$). Subscripts 
denote individual particles and superscripts  
assign a momentum to a particular quantum eigenstate. The two particle momentum eigenstates are normalized such that:
\begin{eqnarray}
\langle \mathbf{p}^{\prime} \mathbf{P}^{\prime} \mid \mathbf{p} \mathbf{P} \rangle = \delta^{3}(\mathbf{p}^{\prime} - \mathbf{p}) \delta^{3}(\mathbf{P}^{\prime} - \mathbf{P}) \label{norm1}, \\ 
 \int \,d^{3}\mathbf{p} \,d^{3}\mathbf{P} \mid \mathbf{p} \mathbf{P} \rangle \langle \mathbf{p} \mathbf{P} \mid = {\rm 1 \hspace{-0.6ex} I} 
 \label{norm2}
\end{eqnarray}
and the transition from the total and relative to the individual momenta can be achieved using:
\begin{eqnarray}
\mathbf{p}_{1} = \mathbf{p} + \frac{1}{2} \mathbf{P} \nonumber \\
\mathbf{p}_{2} = \frac{1}{2} \mathbf{P} - \mathbf{p} \label{trans2},
\end{eqnarray} 
where in (\ref{trans2}) and in the following we assume that the difference between the proton and neutron mass is negligible. 

We examine the case of deuteron disintegration ($e + ^{2}\!\mathrm{H} \rightarrow e + p + n$) where the 2N system is treated in the non-relativistic approximation. In the initial state the deuteron is at rest ($\mathbf{P} = 0$) and the electron has a momentum magnitude of $q_{e}$. We assume that the rest mass of the electron is negligible in comparison to its kinetic energy therefore the initial electron energy $E_{e} \approx q_{e}$. 
The final electron momentum magnitude is $q_{e}^{\prime}$, the final energy $E_{e}^{\prime} \approx q_{e}^{\prime}$ and the electron scattering angle is $\theta_{e}$ 
The magnitude of the three momentum transferred to the 2N system is 
\begin{equation}
Q = \sqrt{q_{e}^{2} + q_{e}^{\prime 2} - 2 q_{e} q_{e}^{\prime} \cos{\theta_{e}}} \approx \sqrt{E_{e}^{2} + E_{e}^{\prime 2} - 2 E_{e} E_{e}^{\prime} \cos{\theta_{e}}}.
\label{momentumQ}
\end{equation}
We work in a reference frame, where the momentum transfer is parallel to $\hat{\mathbf{z}}$. In this frame momentum conservation leads to the expression for the total momentum of the proton and the neutron in the final state:
\begin{equation}
	\mathbf{P^{\mathrm{f}}} = \mathbf{p}_{\mathrm{1}} + \mathbf{p}_{\mathrm{2}} = (0 , 0 , Q).
\end{equation}
The magnitude of the final relative momentum can be calculated from the energy conservation:
\begin{equation}
	|\mathbf{p^{\mathrm{f}}}| = |\frac{1}{2}(\mathbf{p}_{\mathrm{1}} - \mathbf{p}_{\mathrm{2}})| = \frac{1}{2} \sqrt{4 m (E_{d} + E_{e} - E_{e}^{\prime}) - Q^{2}},
	\label{relmom}
\end{equation}
where $E_{d}$ is the (negative) deuteron binding energy and the direction of $\mathbf{p^{\mathrm{f}}}$ can be 
arbitrary. The crucial nuclear matrix element $\mathbf{M}^{\mu}$ between the initial deuteron state (where the total momentum $\mathbf{P} = 0$ and the two particle total angular momentum has a $\hat{z}$ projection $m_{d}$) and the final 2N scattering state can be expressed in terms of the full 2N current operator ($j^{\mu}_{2N}$) and the $t$ operator:
\begin{eqnarray}
\mathbf{M}^{\mu}\left(\mathbf{p}^{\mathrm{f}} , \mathbf{P}^{\mathrm{f}}\right) \nonumber \\
\equiv _{a}\langle  \mathbf{p}^{\mathrm{f}} \mathbf{P}^{\mathrm{f}} , m_{1} \nu_{1} , m_{2} \nu_{2} \mid ({\rm 1 \hspace{-0.6ex} I} + t(E) G_{0}(E)) j^{\mu}_{2N} \mid \phi_{d} \, m_{d} \, \mathbf{P}=0 \rangle   \nonumber \\
= _{a}\langle  \mathbf{p}^{\mathrm{f}} \mathbf{P}^{\mathrm{f}} , m_{1} \nu_{1} , m_{2} \nu_{2} \mid ({\rm 1 \hspace{-0.6ex} I} + t(E) G_{0}(E))\nonumber \\
 (j^{\mu}(1) + j^{\mu}(2) + j^{\mu}(1 , 2)) \mid \phi_{d} \, m_{d} \, \mathbf{P}=0 \rangle   \nonumber \\
= 2 _{a}\langle  \mathbf{p}^{\mathrm{f}} \mathbf{P}^{\mathrm{f}} , m_{1} \nu_{1} , m_{2} \nu_{2} \mid  j^{\mu}(2) \mid \phi_{d} \, m_{d} \, \mathbf{P}=0 \rangle   \nonumber \\
+ _{a}\langle  \mathbf{p}^{\mathrm{f}} \mathbf{P}^{\mathrm{f}}  , m_{1} \nu_{1} , m_{2} \nu_{2} \mid  j^{\mu}(1,2) \mid \phi_{d} \, m_{d} \, \mathbf{P}=0 \rangle   \nonumber \\
+ _{a}\langle  \mathbf{p}^{\mathrm{f}} \mathbf{P}^{\mathrm{f}}  , m_{1} \nu_{1} , m_{2} \nu_{2} \mid t G_{0} j^{\mu}_{2N} \mid \phi_{d} \, m_{d} \, \mathbf{P}=0 \rangle   \label{tG0j2} \label{tG0j12}.
\end{eqnarray}
In this equation the final state is anti-symmetrized:
\begin{equation}
_{a}\langle \mathbf{p} \mathbf{P} , m_{1} \nu_{1} , m_{2} \nu_{2} \mid \equiv \frac{1}{2} \left( \langle \mathbf{p} \mathbf{P} , m_{1} \nu_{1} , m_{2} \nu_{2} \mid - \langle -\mathbf{p} \mathbf{P} , m_{2} \nu_{2} , m_{1} \nu_{1} \mid \right) \label{anti}.
\end{equation}
In the first term on the right hand side of (\ref{anti}) particle $1$ has a spin (isospin) $\hat{\mathbf{z}}$ projection $m_{1}$ ($\nu_{1}$), particle $2$ has a spin (isospin) $\hat{\mathbf{z}}$ projection $m_{2}$ ($\nu_{2}$) and in the second term on the right hand side of (\ref{anti}) the particles are exchanged. The $j(1)$ $\left(j(2)\right)$ operator is a single nucleon current acting on the degrees of freedom of the first (second) nucleon. The $j(1 , 2)$ operator accounts for processes where two nucleons are involved, $t$ is the 2N transition operator and $G_{0}$ is the free 2N propagator. The energy argument of the transition operator and the propagator is
$E = (\mathbf{p}^{\mathrm{f}})^{2}/m$.
Finally, the $\mu$ index denotes the component of the current operator. In particular $\mu = 0$ stands for charge density operators while $\mu = 1,2,3$ stand for spatial components. In the  following sections the way we calculate the individual parts of $\mathbf{M}^{\mu}$ in (\ref{tG0j12}) will be discussed separately for a specific choice of the coordinate system and value of $\mu$; we will drop $\mu$ for brevity.

\section{Deuteron bound state}

The structure of the deuteron wave function can be written in the operator form, following \cite{deuteronspin1,fach01,deuteronscalar}:
\begin{flalign}
	& \mid \phi_{d} \, m_{d} \, \mathbf{P}=0 \rangle    \nonumber = & \\
	& =\int d^{3}\mathbf{p} \mid \mathbf{p} \mathbf{P}=0 \rangle\sum_{l = 1}^{2} \phi_{l}(|\mathbf{p}|) \left[{\rm 1 \hspace{-0.6ex} I}^{\mathrm{isospin}} \otimes b_{l}(\mathbf{p})^{\mathrm{spin}} \right] \left[\mid 0 \, 0 \rangle \otimes \chi(m_{d}) \right] \nonumber & \\
	& \equiv \int d^{3}\mathbf{p} \mid \mathbf{p} \mathbf{P}=0 \rangle\sum_{l = 1}^{2} \phi_{l}(|\mathbf{p}|) \left[B_{l}(\mathbf{p}) \right] \left[\mid 0 \, 0 \rangle \otimes \chi(m_{d}) \right]	. &\label{deuteron}
\end{flalign}
where
\[
	b_{1} = {\rm 1 \hspace{-0.6ex} I}
\]
\[
	b_{2} = \left( \boldsymbol{\sigma}(1) \cdot \mathbf{p} \boldsymbol{\sigma}(2) \cdot \mathbf{p}  - \frac{1}{3} \mathbf{p} \cdot \mathbf{p} {\rm 1 \hspace{-0.6ex} I} \right).
\]
In equation (\ref{deuteron}) $\mid \chi(m_{d})\rangle$ is a state in which the spins of the two spin $\frac{1}{2}$ particles are coupled to a total spin $1$ with a $\hat{\mathbf{z}}$ projection $m_{d}$. The isospins of the two nucleons are coupled to the total isospin $0$ state $\mid 0 \, 0 \rangle$. $\phi_{l}$ are scalar functions of the relative momentum and $\boldsymbol{\sigma}(1)$, $\boldsymbol{\sigma}(2)$ are doubled spin operators in the spin space of one nucleon and identity operators in the spin space of the other nucleon, respectively:
\begin{equation}
	\boldsymbol{\sigma}(1) = \left( \sigma^{x} \otimes {\rm 1 \hspace{-0.6ex} I} , \sigma^{y} \otimes {\rm 1 \hspace{-0.6ex} I} , \sigma^{z} \otimes {\rm 1 \hspace{-0.6ex} I} \right) ,
\label{sigma1}
\end{equation}
\begin{equation}
	\boldsymbol{\sigma}(2) = \left( {\rm 1 \hspace{-0.6ex} I} \otimes \sigma^{x} , {\rm 1 \hspace{-0.6ex} I} \otimes \sigma^{y} , {\rm 1 \hspace{-0.6ex} I} \otimes \sigma^{z} \right).
\label{sigma2}
\end{equation} 
Vector components in (\ref{sigma1}) and (\ref{sigma2}) are operators written in terms of the tensor product $\otimes$. They act in the $4$ dimensional spin space of the two nucleon system and can be represented by $4 \times 4$ matrices - tensor products of identity operators and Pauli matrices.

Note that in equation (\ref{deuteron}) states and operators in the 2N isospin - spin space are placed inside the $\left[ \dots \right]$ brackets; in Sec. \ref{sec:representation} we will introduce a simple way to implement these expressions as $16$ dimensional vectors and $16 \times 16$ matrices for use in our numerical treatment.

Scalar functions $\phi_{l}$ in expansion (\ref{deuteron}) can be calculated using three dimensional formalism, see for example \cite{deuteronscalar}. Nowadays deuteron bound state calculations can use any 2N potentials given in the operator form and do not require substantial computational resources. 

\section{Single nucleon currents in three dimensions.}

Single nucleon (1N) currents act on the degrees of freedom of one particle. Their matrix elements in the momentum space depend only on the initial and final momenta
and are operators in the isospin - spin space. For example the matrix element for the second nucleon, $j(2)$, reads:
\begin{equation}
\left[ \langle \mathbf{p}_{1}^{\prime} \mathbf{p}_{2}^{\prime} \mid j(2) \mid \mathbf{p}_{1} \mathbf{p}_{2} \rangle \right]= \delta \left( \mathbf{p}^{\prime}_{1} - \mathbf{p}_{1} \right) \left[j(2 , \mathbf{p}^{\prime}_{2} - \mathbf{p}_{2} , \mathbf{p}^{\prime}_{2} + \mathbf{p}_{2})\right] 
\end{equation}
where in view of the standard nonrelativistic current, the dependence on the difference and sum of the initial and final momenta is used. Implementing the transition from the individual particle momenta to the relative momenta leads to
\begin{eqnarray}
\left[ \langle \mathbf{p}^{\prime} \mathbf{P}^{\prime} \mid j(2) \mid \mathbf{p} \mathbf{P} \rangle \right] = \delta \left( \frac{1}{2} \mathbf{P}^{\prime} - \frac{1}{2} \mathbf{P} + \mathbf{p}^{\prime} - \mathbf{p} \right) \nonumber \\
\left[j(2 ,  \frac{1}{2} \mathbf{P}^{\prime} - \frac{1}{2} \mathbf{P} - \mathbf{p}^{\prime} + \mathbf{p} ,  \frac{1}{2} \mathbf{P}^{\prime} + \frac{1}{2} \mathbf{P} - \mathbf{p}^{\prime} - \mathbf{p})\right].
\end{eqnarray}
Again, the expressions inside the square brackets $\left[ \dots \right]$ can be easily represented using the notion of the Kronecker product, see
Sec. \ref{sec:representation}. The action of $j(2)$ on the deuteron state at rest can be worked out:
\begin{flalign}
&\left[ \langle \mathbf{p}^{\prime} \mathbf{P}^{\prime} \mid j(2)\mid \phi_{d} \, m_{d} \, \mathbf{P}=0  \rangle \right] = \nonumber & \\
&= \sum_{l = 1}^{2} \phi_{l}(|\mathbf{p}^{\prime} + \frac{1}{2} \mathbf{P}^{\prime}|) \left[j(2 , \mathbf{P}^{\prime} , -2 \mathbf{p}^{\prime})\right] \left[ B_{l}(\mathbf{p}^{\prime} + \frac{1}{2} \mathbf{P}^{\prime}) \right] \left[\mid 0 \, 0 \rangle \otimes \chi(m_{d}) \right] \nonumber &\\
&\equiv \left[ O^{\mathrm{1N}}(2 , \mathbf{p}^{\prime} , \mathbf{P}^{\prime}) \right] \left[\mid 0 \, 0 \rangle \otimes \chi(m_{d}) \right] , &\label{j1N}
\end{flalign}
where we used
$\mathbf{P} = 0$ and the normalization of momentum eigenstates (\ref{norm1})-(\ref{norm2}) so $O^{\mathrm{1N}}$ is the resulting single particle operator. Equation (\ref{j1N}) gives the full isospin - spin state for the final
$\mathbf{p}^{\prime}$, $\mathbf{P}^{\prime}$ momenta. 

\section{2N currents in three dimensions.}

For a wide class of 2N current operators, their matrix elements in the momentum space (operators in isospin - spin space) are given in the form:
\begin{equation}
\left[ \langle \mathbf{p}_{1}^{\prime} \mathbf{p}_{2}^{\prime} \mid j(1,2) \mid \mathbf{p}_{1} \mathbf{p}_{2} \rangle \right] = \left[j(1 , 2 , \mathbf{p}^{\prime}_{1} - \mathbf{p}_{1} , \mathbf{p}^{\prime}_{2} - \mathbf{p}_{2})\right] ,
\label{2Nform}
\end{equation}
see for example \cite{PSC31471,PRC402294,PRC41309}. The right hand side of (\ref{2Nform}) is as a linear combination of scalar functions ($f_{i}^{jS}$ , $f_{i}^{j}$) and products of spin space operators ($O_{jS}$ , $\mathbf{O}_{j}$) and isospin space operators ($T_{i}$):
\begin{eqnarray}
\left[ j^{0}(1,2) \right] = \sum_{i = 1}^{5} \sum_{j = 1}^{8} f_{i}^{jS}(\mathbf{p}^{\prime}_{1} - \mathbf{p}_{1} , \mathbf{p}^{\prime}_{2} - \mathbf{p}_{2}) \left[ T_{i} O_{jS} \right].
\label{2Ndecomp2} \\
\left[ \vec{j}(1,2) \right] = \sum_{i = 1}^{5} \sum_{j = 1}^{24} f_{i}^{j}(\mathbf{p}^{\prime}_{1} - \mathbf{p}_{1} , \mathbf{p}^{\prime}_{2} - \mathbf{p}_{2}) \left[ T_{i} \mathbf{O}_{j} \right], \label{2Ndecomp1}
\end{eqnarray}
where the subscript S distinguishes density and current components. In Ref. \cite{2Ncurrents1} a general operator basis for the local 2N current operator was introduced.
In Sec. \ref{sec:representation} some examples of operators from (\ref{2Ndecomp2}), (\ref{2Ndecomp1}) will be used to demonstrate our matrix representation of expressions inside $\left[ \dots \right]$.

Again, using (\ref{trans2}), current matrix elements become:
\begin{equation}
\left[ \langle \mathbf{p}^{\prime} \mathbf{P}^{\prime} \mid j(1,2) \mid \mathbf{p} \mathbf{P} \rangle \right] = \left[j(1 , 2 , \frac{1}{2} \mathbf{P}^{\prime} - \frac{1}{2} \mathbf{P} + \mathbf{p}^{\prime} - \mathbf{p} , \frac{1}{2} \mathbf{P}^{\prime} - \frac{1}{2} \mathbf{P} - \mathbf{p}^{\prime} + \mathbf{p})\right] 
\end{equation}
In the following we restrict ourselves 
to this class of momentum dependences. Our approach can, however, be generalized to include any type of momentum dependence. The action of $j(1 , 2)$ on the deuteron state can be worked out using (\ref{norm1}),(\ref{norm2}), (\ref{trans2}) and 
$\mathbf{P} = 0$. In the laboratory frame it yields:
\begin{eqnarray}
\left[ \langle \mathbf{p}^{\prime} \mathbf{P}^{\prime} \mid j(1 , 2) \mid \phi_{d} \, m_{d} \, \mathbf{P}=0 \rangle \right]  \nonumber \\
= \int \,d^{3}\mathbf{p}^{\prime\prime} \sum_{l = 1}^{2} \phi_{l}(|\mathbf{p}^{\prime\prime}|) \left[j(1 , 2 , \frac{1}{2} \mathbf{P}^{\prime} + \mathbf{p}^{\prime} - \mathbf{p}^{\prime\prime} , \frac{1}{2} \mathbf{P}^{\prime} - \mathbf{p}^{\prime} + \mathbf{p}^{\prime\prime})\right] \nonumber \\ \left[ B_{l}(\mathbf{p}^{\prime\prime}) \right] \left[\mid 0 \, 0 \rangle \otimes \chi(m_{d}) \right] \nonumber \\
 \equiv \left[ O^{\mathrm{2N}}(1 , 2 , \mathbf{p}^{\prime} , \mathbf{P}^{\prime}) \right] \left[\mid 0 \, 0 \rangle \otimes \chi(m_{d}) \right]
 \label{j2N} ,
\end{eqnarray}
where $O^{\mathrm{2N}}$ is the resulting two-particle operator. 
Equation (\ref{j2N}) gives the full isospin - spin state for the final
$\mathbf{p}^{\prime}$, $\mathbf{P}^{\prime}$ momenta. 

\section{$t$ operator in three dimensions}

The $t$ operator satisfies the Lippmann - Schwinger equation:
\begin{equation}
t(E) = V + t(E) G_{0}(E) V
\label{LSEleft}
\end{equation}
or equivalently 
\begin{equation}
t(E) = V + V G_{0}(E) t(E)
\label{LSEright}
\end{equation}
where $G_{0}(E)$ is the free propagator depending on the energy $E$
and $V$ is a 2N potential. It follows that, as shown in \cite{decomp} also $t$ can be written as a linear combination of scalar functions ($t_{\gamma , i}$) and operators ($\mathbf{W}_{\gamma , i}$) in the isospin - spin space:
\begin{equation}
\left[ \langle \mathbf{p}^{\prime} \mid t(E) \mid \mathbf{p} \rangle \right] = \sum_{\gamma} \sum_{i = 1}^{6} t_{\gamma , i}(|\mathbf{p}^{\prime}| , |\mathbf{p}| , \hat{\mathbf{p}}^{\prime} \cdot \hat{\mathbf{p}} , E) \left[\mathbf{W}_{\gamma , i}(\hat{\mathbf{p}}^{\prime} , \hat{\mathbf{p}}) \right].
\label{decomp1}
\end{equation}
Here
\begin{equation} 
\left[\mathbf{W}_{\gamma , i}(\hat{\mathbf{p}}^{\prime} , \hat{\mathbf{p}}) \right] = \left[\mathbf{C}_{\gamma}^{\mathrm{isospin}} \otimes \mathbf{w}_{i}^{\mathrm{spin}}(\mathbf{p}^{\prime} , \mathbf{p})\right] \nonumber
\end{equation}
are again operators in the isospin - spin space (matrix elements between momentum states), with $\mathbf{w}_{i}$ ($i = 1,2,...,6$) acting in the $4$ dimensional spin space of the 2N system.
The decomposition (\ref{decomp1}) 
is not unique; our choice of the six $\mathbf{w}_{i}$ operators is consistent with \cite{deuteronscalar}
. Scalar functions arising in the decomposition of $t$ can be calculated in the three dimensional formalism. Calculations can be performed for any type of the NN potential satisfying a similar expansion (\ref{decomp1}).
For details see Ref. \cite{deuteronscalar}.
\begin{eqnarray}
\mathbf{w}_{1}(\mathbf{p}^{\prime} , \mathbf{p}) = {\rm 1 \hspace{-0.6ex} I} ,\\
\mathbf{w}_{2}(\mathbf{p}^{\prime} , \mathbf{p}) = \boldsymbol{\sigma}(1) \cdot \boldsymbol{\sigma}(2) ,\\
\mathbf{w}_{3}(\mathbf{p}^{\prime} , \mathbf{p}) = i (\boldsymbol{\sigma}(1) + \boldsymbol{\sigma}(2)) \cdot (\hat{\mathbf{p}} \times \hat{\mathbf{p}}^{\prime}) ,\\
\mathbf{w}_{4}(\mathbf{p}^{\prime} , \mathbf{p}) = \boldsymbol{\sigma}(1) \cdot (\hat{\mathbf{p}} \times \hat{\mathbf{p}}^{\prime}) \, \boldsymbol{\sigma}(2) \cdot (\hat{\mathbf{p}} \times \hat{\mathbf{p}}^{\prime}) ,\\
\mathbf{w}_{5}(\mathbf{p}^{\prime} , \mathbf{p}) = \boldsymbol{\sigma}(1) \cdot (\hat{\mathbf{p}}^{\prime} + \hat{\mathbf{p}}) \, \boldsymbol{\sigma}(2) \cdot (\hat{\mathbf{p}}^{\prime} + \hat{\mathbf{p}}) ,\\
\mathbf{w}_{6}(\mathbf{p}^{\prime} , \mathbf{p}) = \boldsymbol{\sigma}(1) \cdot (\hat{\mathbf{p}}^{\prime} - \hat{\mathbf{p}}) \, \boldsymbol{\sigma}(2) \cdot (\hat{\mathbf{p}}^{\prime} - \hat{\mathbf{p}})
\end{eqnarray}
The $C_{\gamma} = \mid \gamma \rangle \langle \gamma \mid$ isospin operators project onto one of the four 2N isospin states:
\begin{equation}
	\mid \gamma \rangle = \mid \left( \frac{1}{2} \, \frac{1}{2} \right) t=0 , 1 \, m_{t} = -t \dots t \rangle.
\end{equation}
The $\mid \gamma \rangle$ states are chosen in this way, because $t(E)$ conserves the total 2N isospin.

The rescattering part of the matrix element $M$ in (\ref{tG0j12}) can be written as: 
\begin{eqnarray}
\left[ \langle \mathbf{p}^{\prime} \mathbf{P}^{\prime} \mid t(E) G_{0} j_{2N} \mid \phi_{d} \, m_{d} \, \mathbf{P}=0 \rangle \right]  \nonumber \\
= \int \,d^{3}\mathbf{p} \left[ \langle \mathbf{p}^{\prime} \mid t(E) \mid \mathbf{p} \rangle \right] \frac{1}{E - \frac{\mathbf{p}^{2}}{m} + i \epsilon} \nonumber \\ 
\times \left[ O(\mathbf{p} , \mathbf{P}^{\prime}) \right] \left[\mid 0 \, 0 \rangle \otimes \chi(m_{d}) \right] \nonumber\\
=  m \int_{0}^{\bar{p}}  \frac{\mathbf{p}^{2} \left[ \mathbf{f}(|\mathbf{p}|) \right] - \mathbf{p}^{\prime 2} \left[ \mathbf{f}(|\mathbf{p}^{\prime}|) \right]}{\mathbf{p}^{\prime 2} - \mathbf{p}^{2}} \, d|\mathbf{p}| \nonumber \\
+ m \frac{|\mathbf{p}^{\prime}|  \left[ \mathbf{f}(|\mathbf{p}^{\prime}|) \right]}{2} \left( \ln \left( \frac{\bar{p} + |\mathbf{p}^{\prime}|}{\bar{p} - |\mathbf{p}^{\prime}|}\right)  - i \pi \right)  \nonumber \\
\times \left[\mid 0 \, 0 \rangle \otimes \chi(m_{d}) \right]
\label{tG0j}
\end{eqnarray}
where $O$ is either $O^{1N}$ from Eq. (\ref{j1N}) or $O^{2N}$ from Eq. (\ref{j2N}), $E = \mathbf{p}^{\mathrm{\prime 2}}/m$ is the relative energy of the final 2N state and
\begin{equation}
\left[ \mathbf{f}(|\mathbf{p}|) \right] = \int_{0}^{2 \pi} d\phi \int_{-1}^{1} dx \left[ \langle  \mathbf{p}^{\prime} \mid t(E) \mid \mathbf{p} \rangle \right] \left[ O(\mathbf{p} , \mathbf{P}^{\prime}) \right],
\end{equation}
since
\begin{equation}
\mathbf{p} = |\mathbf{p}| (\sqrt{1 - x^{2}} \cos{\phi} , \sqrt{1 - x^{2}} \sin{\phi} , x).
\end{equation}

The integral (\ref{tG0j}) with the cut-off value $\bar{p}$ can be easily calculated numerically. In the next section we show how to prepare its component $\left[ O(\mathbf{p} , \mathbf{P}^{\prime}) \right]$.

\section{Representation of spin-isospin operators}

\label{sec:representation}

Now that the form of expressions in (\ref{tG0j12}) has been established it remains to find a way to represent operators and states inside $\left[ \dots \right]$. Once an appropriate matrix representation is found, numerical calculations can be constructed using equations (\ref{j1N}), (\ref{j2N}) and (\ref{tG0j}) by simple substitutions and matrix multiplications. 

Our choice for the $16$ dimensional basis of the two nucleon isospin - spin state space (the deuteron in the initial state, the proton and the neutron in the final state) is the set of vector states $\{ \mid e_{i} \rangle \}$ ($i = 1,2,...,16$):
\begin{equation}
\mid e_{i} \rangle  = \left( \mid m_{1}^{\mathrm{isospin}}(i) \rangle \otimes \mid m_{2}^{\mathrm{isospin}}(i)\rangle \right) \otimes \left( \mid m_{1}^{\mathrm{spin}}(i) \rangle \otimes \mid m_{2}^{\mathrm{spin}}(i) \rangle \right)
\end{equation}
where $m^{\mathrm{spin(isospin)}}_{j}$ are the spin (isospin) projections of nucleon 
$j$ and the corresponding quantum numbers are given in Tab. \ref{fullstates}
\begin{center}
\begin{table}\centering
 \setlength{\extrarowheight}{4pt}
 \caption{Reference quantum numbers for our KP states.}
\begin{tabular}{c|cccc}
 $i$ & $m_{1}^{\mathrm{isospin}}(i)$ & $m_{2}^{\mathrm{isospin}}(i)$ & $m_{1}^{\mathrm{spin}}(i)$ & $m_{2}^{\mathrm{spin}}(i)$ \\ \hline
 $1$ & $\frac{1}{2}$ & $\frac{1}{2}$ & $\frac{1}{2}$ & $\frac{1}{2}$  \\
 $2$ & $\frac{1}{2}$ & $\frac{1}{2}$ & $\frac{1}{2}$ & $-\frac{1}{2}$  \\
 $3$ & $\frac{1}{2}$ & $-\frac{1}{2}$ & $\frac{1}{2}$ & $\frac{1}{2}$  \\
 $4$ & $\frac{1}{2}$ & $-\frac{1}{2}$ & $\frac{1}{2}$ & $-\frac{1}{2}$  \\
 $5$ & $\frac{1}{2}$ & $\frac{1}{2}$ & $-\frac{1}{2}$ & $\frac{1}{2}$  \\
 $6$ & $\frac{1}{2}$ & $\frac{1}{2}$ & $-\frac{1}{2}$ & $-\frac{1}{2}$  \\
 $7$ & $\frac{1}{2}$ & $-\frac{1}{2}$ & $-\frac{1}{2}$ & $\frac{1}{2}$  \\
 $8$ & $\frac{1}{2}$ & $-\frac{1}{2}$ & $-\frac{1}{2}$ & $-\frac{1}{2}$  \\
 $9$ & $-\frac{1}{2}$ & $\frac{1}{2}$ & $\frac{1}{2}$ & $\frac{1}{2}$  \\
 $10$ & $-\frac{1}{2}$ & $\frac{1}{2}$ & $\frac{1}{2}$ & $-\frac{1}{2}$  \\
 $11$ & $-\frac{1}{2}$ & $-\frac{1}{2}$ & $\frac{1}{2}$ & $\frac{1}{2}$  \\
 $12$ & $-\frac{1}{2}$ & $-\frac{1}{2}$ & $\frac{1}{2}$ & $-\frac{1}{2}$  \\
 $12$ & $-\frac{1}{2}$ & $\frac{1}{2}$ & $-\frac{1}{2}$ & $\frac{1}{2}$  \\
 $14$ & $-\frac{1}{2}$ & $\frac{1}{2}$ & $-\frac{1}{2}$ & $-\frac{1}{2}$  \\
 $15$ & $-\frac{1}{2}$ & $-\frac{1}{2}$ & $-\frac{1}{2}$ & $\frac{1}{2}$  \\
 $16$ & $-\frac{1}{2}$ & $-\frac{1}{2}$ & $-\frac{1}{2}$ & $-\frac{1}{2}$ .
  \label{fullstates}
\end{tabular}
\end{table}
\end{center}
Any operator or state
in this space can be constructed using the notion of the Kronecker product (KP). The Mathematica $^{\mbox{\textregistered}}$ \cite{math} symbolic programming software contains definitions for the KP, which makes translating expressions for isospin - spin operators a simple task. However, it is important to remember that the order of operators in the KP must be preserved. Tab. \ref{fullstates} can serve as a reference to keep consistence with this paper.  

The deuteron operators $[B_{1}(\mathbf{p})]$ and $[B_{2}(\mathbf{p})]$ with $\mathbf{p} = \left(p_{x} , p_{y} , p_{z} \right)$ from equation (\ref{deuteron}) have a form simple enough to have their matrix representation written out in full. $[B_{1}]$ is simply a $16 \times 16$ identity matrix. $[B_{2}]$ has a block diagonal form:
\begin{equation}
\left(
\begin{array}{cccc}
 B & 0 & 0 & 0 \\
 0 & B & 0 & 0 \\
 0 & 0 & B & 0 \\
 0 & 0 & 0 & B
\end{array}
\right)\nonumber
\end{equation}
where $B$ is a $4 \times 4$ matrix:
\begin{tiny}
\begin{equation}
\left(
\begin{array}{cccc}
 \frac{1}{3} \left(-p_x^2-p_y^2-p_z^2\right)+p_z^2 & p_z \left(p_x-i p_y\right) & p_z \left(p_x-i p_y\right) & \left(p_x-i p_y\right){}^2 \\
 p_z \left(p_x+i p_y\right) & \frac{1}{3} \left(-p_x^2-p_y^2-p_z^2\right)-p_z^2 & \left(p_x-i p_y\right) \left(p_x+i p_y\right) & p_z \left(-\left(p_x-i p_y\right)\right) \\
 p_z \left(p_x+i p_y\right) & \left(p_x-i p_y\right) \left(p_x+i p_y\right) & \frac{1}{3} \left(-p_x^2-p_y^2-p_z^2\right)-p_z^2 & p_z \left(-\left(p_x-i p_y\right)\right) \\
 \left(p_x+i p_y\right){}^2 & p_z \left(-\left(p_x+i p_y\right)\right) & p_z \left(-\left(p_x+i p_y\right)\right) & \frac{1}{3} \left(-p_x^2-p_y^2-p_z^2\right)+p_z^2
\end{array}
\right)\nonumber.
\end{equation}
\end{tiny}

Operators appearing in (\ref{2Ndecomp2}) and (\ref{2Ndecomp1}) have a more complicated isospin operator form. A simple example of their matrix representation is given below in Fig. \ref{T2O5mu3}.

\begin{figure}[h]\centering
\includegraphics[width=1.0\textwidth,angle=0]{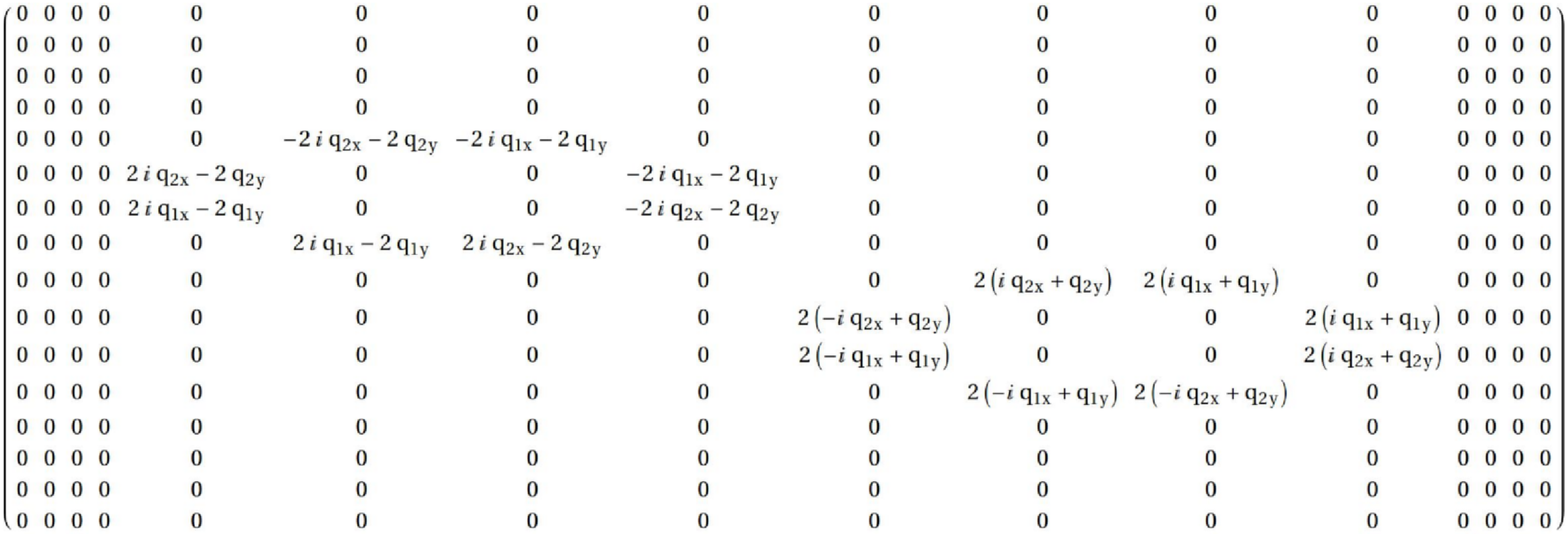}
\caption{$\left[ T_{2} O_{5}^{3} \right] \left( \mathbf{p}_{1}^{\prime} - \mathbf{p}_{1} = \left( q_{1x} , q_{1y} , q_{1z} \right) , \mathbf{p}_{2}^{\prime} - \mathbf{p}_{2} = \left( q_{2x} , q_{2y} , q_{2z} \right) \right)$ with $T_{2} = \boldsymbol{\tau}_{1}^{z}-\boldsymbol{\tau}_{2}^{z}$ and $O_{5} = (\mathbf{q}_{1} \times \boldsymbol{\sigma}_{1})+(\mathbf{q}_{2} \times \boldsymbol{\sigma}_{2})$. $ \boldsymbol{\tau}_{i}$ ($ \boldsymbol{\sigma}_{i}$) is the 2N isospin (spin) vector operator acting in the space of nucleon $i$.}
\label{T2O5mu3}
\end{figure}

States can be constructed in a similar manner using the built in Mathematica $^{\mbox{\textregistered}}$ definitions for Clebsch-Gordan coefficients, for example $\left[\mid 0 \, 0 \rangle \otimes \chi(1) \right]$ results in:
\begin{equation}
\left(0,0,0,0,\frac{1}{\sqrt{2}},0,0,0,-\frac{1}{\sqrt{2}},0,0,0,0,0,0,0\right).
\end{equation}

\section{Results}

In the following we will present results obtained using a chiral NNLO potential \cite{Epelbaum:2004fk} with $\Lambda = 550 $ MeV$/$c and $\tilde{\Lambda} = 600$ MeV$/$c.
The operator form of such a potential was briefly described in Appendix~C of Ref.~\cite{deuteronscalar},
where also an example set of necessary parameters was given for its neutron-proton version. 
The same parameters will also be used in the present paper.

There are three basic ingredients in our calculations: the deuteron wave function,
the 2N t-matrix and the 2N current operator. Before we show selected observables 
for the $ ^2{\rm H} ( e, e' p) n $ reaction, we will describe our numerical performance
and the way we verify the quality of our calculations.

As described in Ref.~\cite{deuteronscalar} and equation (\ref{deuteron}), the deuteron in the operator form
is represented by two functions $\phi_1(p)$ and $\phi_2(p)$.
The corresponding Schr\"odinger equation for $\phi_1(p)$ and $\phi_2(p)$
can be rewritten as an eigenvalue problem,
which is of the same type and dimension as 
the one solved for the
deuteron wave function in the standard partial wave representation,
where one deals with the $s$- and $d$-components, $\psi_0(p) $ and $\psi_2(p)$.
The connection between the solutions is very simple \cite{fach01}
\begin{eqnarray}
\psi_0 (p) = \sqrt{4 \pi} \,\phi_1 (p), \, \, \, \,
\psi_2 (p) = \frac{4 \sqrt{2} \pi p^2 } { 3 } \, \phi_2 (p) 
\end{eqnarray}
and can be used to check the numerical performance.

\begin{figure}[hp]\centering
\includegraphics[width=0.45\textwidth,angle=0]{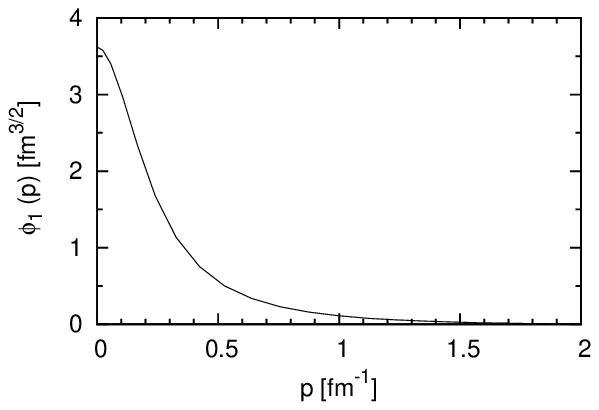}
\includegraphics[width=0.45\textwidth,angle=0]{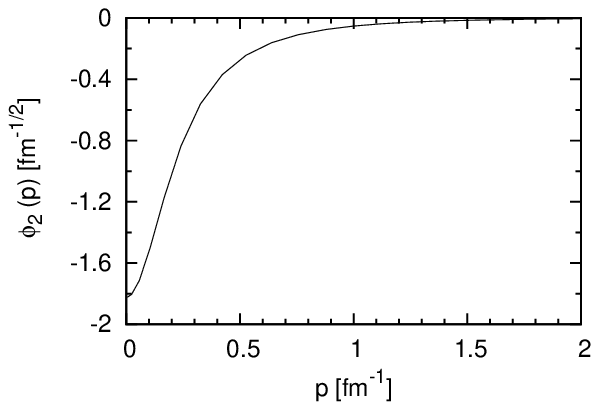}
\caption{The $\phi_1(p)$ (left) and $\phi_2(p)$ (right) expansion
function in the operator form of the deuteron 
as a function of the magnitude of the relative momentum $p$
for the considered chiral NNLO potential. 
}
\label{f1}
\end{figure}

\begin{figure}[hp]\centering
\includegraphics[width=0.45\textwidth,angle=0]{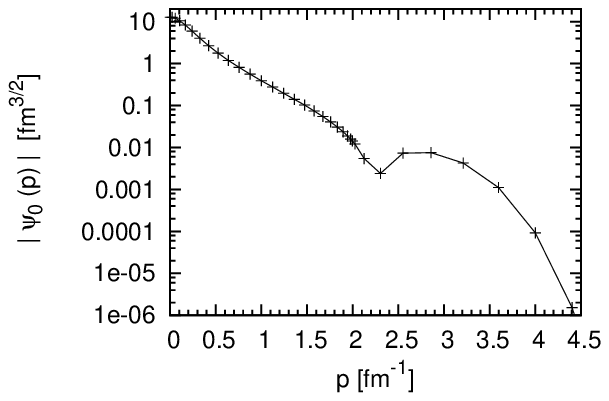}
\includegraphics[width=0.45\textwidth,angle=0]{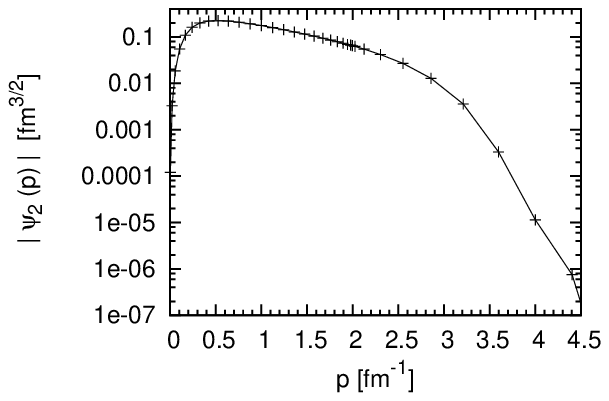}
\caption{The $s$-wave (left) and $d$-wave (right) component of the deuteron
wave function as a function of the magnitude of the relative momentum $p$
for the considered chiral NNLO potential. Crosses
represent results obtained using the
operator approach and solid lines are directly from the standard
partial wave decomposition.}
\label{f2}
\end{figure}

In Fig.~\ref{f1} we show directly the $\phi_1(p)$ and $\phi_2(p)$
functions required for the operator expansion of the deuteron.
In Fig.~\ref{f2} the resulting $s$- and $d$-wave components
in momentum space are compared to the
results obtained by firstly decomposing the NN potential into partial
waves and then solving
the resulting Schr\"odinger equation in its standard form.
The agreement for the two wave function
components is perfect for all their significant values.

In Ref.~\cite{deuteronscalar} we solved the Lippmann-Schwinger equation (LSE) for the 2N t-matrix
directly in three dimensions. At that time we focused mainly on the on-shell
behavior of the expansion coefficients $t_i (p',p,x; E_{2N} ) $, that is 
we were interested in $t_i \left(p_0,p_0,x; E_{2N}=\frac{p_0^2}{m} \right) $, which are sufficient
to calculate the Wolfenstein parameters and the nucleon-nucleon 
scattering observables. Furthermore, we solved the LSE in such a form (Eq.~(2.6) from Ref.~\cite{deuteronscalar})
\begin{eqnarray}
\lefteqn{\sum_{j} A_{ kj} ( {\vec p'}, {\vec p}) t_j^{t m_t}({\vec p'}, {\vec p}) = 
 \sum_{j} A_{ kj} ( {\vec p'}, {\vec p}) v_j^ { t m_t}( {\vec p'}, {\vec p} )} \cr
&+&  \int d^ 3 p'' \sum_{ j j'} v_j^ { t m_t}( {\vec p'}, {\vec p''}) G_0 ( p'')
t_{ j'}^ { t m_t}( {\vec p''}, {\vec p)} B_{ kjj'} ( {\vec p'}, {\vec p''},  {\vec p}),
\label{lse1}
\end{eqnarray} 
that the magnitude of the initial $ {\vec p} $ momentum could be fixed.

Clearly, for the $^2{\rm H} ( e, e' p) n $ reaction we need 
a "left" version of Eq.~(\ref{lse1}), which allows us to find the
half-shell t-matrix for a fixed final relative momentum,
$\vec p^{\mathrm{f}}$, given now by the reaction kinematics. 
The starting point for this new version is equation (\ref{LSEleft}).
Repeating the algebra outlined in Ref.~\cite{deuteronscalar}, we prepared 
a numerical realization of this "left" version of the LSE, leading to
the scalar expansion coefficients 
$t_i \left(\vec p^{\mathrm{f}}, \vec p ; E_{2N}= \frac{ (\mathbf{p}^{\mathrm{f}})^2 }{m}  \right) 
\equiv t_i \left( p^{\mathrm{f}}, p , x ; E_{2N}= \frac{ (\mathbf{p}^{\mathrm{f}})^2 }{m}  \right) $,
where $ x \equiv {\hat p}^{\mathrm{f}} \cdot {\hat p} $. Our numerical 
scheme was again based on matrix inversion and used 
the standard LU decomposition of Numerical Recipes \cite{numrec}.
In order to achieve a unique and smooth solution also for 
$ p = p^{\mathrm{f}}$, it was sufficient to calculate the average
\begin{flalign}
& t_i \left(p^{\mathrm{f}}, p , x ; E_{2N}= \frac{ (\mathbf{p}^{\mathrm{f}})^2 }{m}  \right) = \nonumber & \\
& =
\frac12 \left(
t_i \left(p^{\mathrm{f}} - \delta_{p^{\mathrm{f}}},  p , x ; E_{2N}= \frac{ (\mathbf{p}^{\mathrm{f}})^2 }{m}  \right)
\, + \,
t_i \left(p^{\mathrm{f}} + \delta_{p^{\mathrm{f}}} , p , x ; E_{2N}= \frac{ (\mathbf{p}^{\mathrm{f}})^2 }{m}  \right)
\right) \, , &
\label{lse3}
\end{flalign}
with $ \delta_{p^{\mathrm{f}}} \approx \, $ 0.01 fm$^{-1}$. 

Actually, this effort turned out to be unnecessary and provided 
merely an additional check of numerics, since
\begin{eqnarray}
t_i \left(p^{\mathrm{f}}, p , x ; E_{2N}= \frac{ (\mathbf{p}^{\mathrm{f}})^2 }{m}  \right)
=
t_i \left(p , p^{\mathrm{f}} , x ; E_{2N}= \frac{ (\mathbf{p}^{\mathrm{f}})^2 }{m}  \right) 
\label{lse4}
\end{eqnarray}
for the most general rotational, parity and time
reversal invariant form of the NN force. That means that 
the left coefficients, $ t_i \left(p^{\mathrm{f}} , p , x ; E_{2N}= \frac{ (\mathbf{p}^{\mathrm{f}})^2 }{m}  \right)$,
can be obtained directly from the "right" version of LSE.

\begin{figure}[hp]\centering
\includegraphics[width=0.45\textwidth,angle=0]{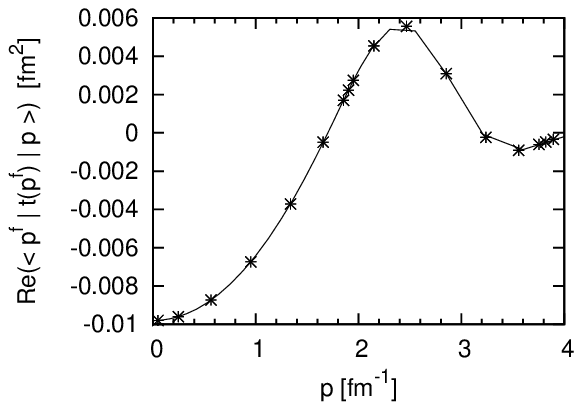}
\includegraphics[width=0.45\textwidth,angle=0]{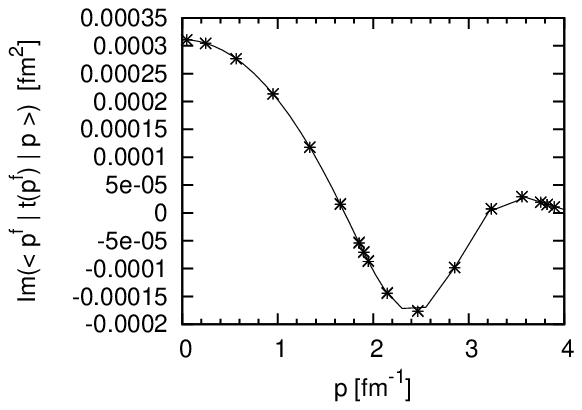}
\caption{The real (left) and imaginary (right) parts of the
half-shell $^1S_0$ t-matrix as a function of the 
initial momentum $p$ for $p^{\mathrm{f}} \approx$ 1.9 fm$^{-1}$. 
Points represent predictions obtained by a projection 
from the three-dimensional results. Solid lines represent 
direct solutions of LSE in the standard partial wave decomposition.}
\label{f3}
\end{figure}

\begin{figure}[hp]\centering
\includegraphics[width=0.45\textwidth,angle=0]{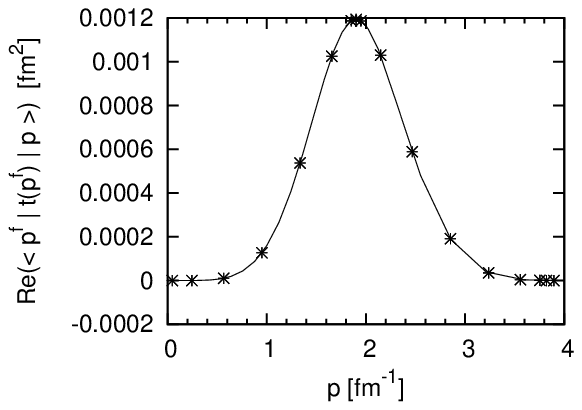}
\includegraphics[width=0.45\textwidth,angle=0]{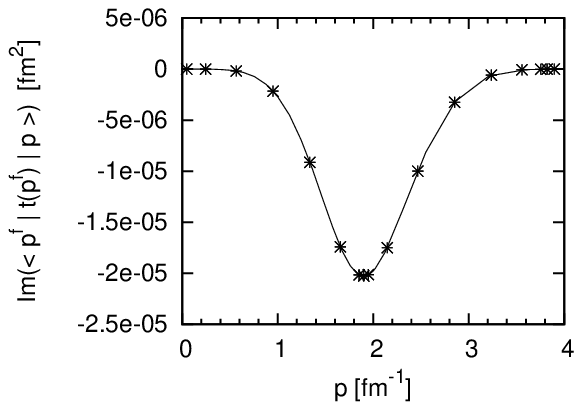}
\caption{The same as in Fig.~\ref{f3} 
for the half-shell $^3H_5$ t-matrix.}
\label{f4}
\end{figure}

\begin{figure}[hp]\centering
\includegraphics[width=0.45\textwidth,angle=0]{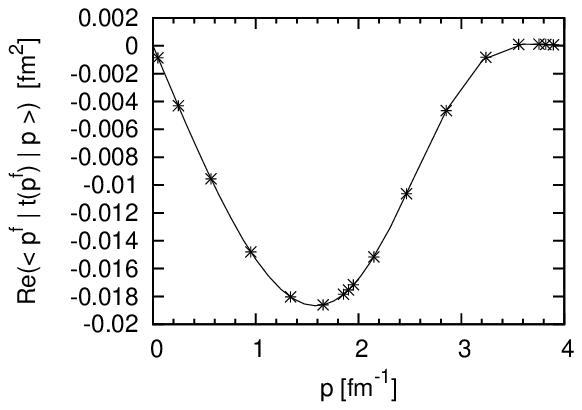}
\includegraphics[width=0.45\textwidth,angle=0]{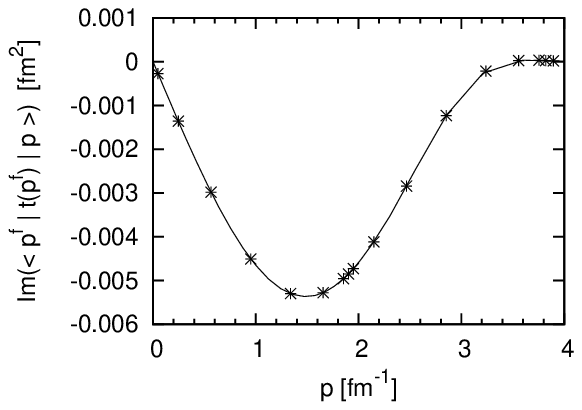}
\includegraphics[width=0.45\textwidth,angle=0]{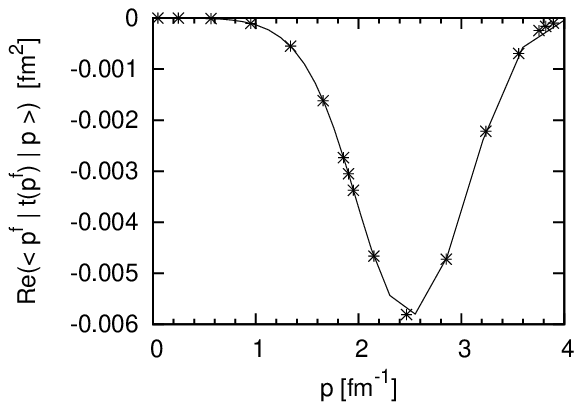}
\includegraphics[width=0.45\textwidth,angle=0]{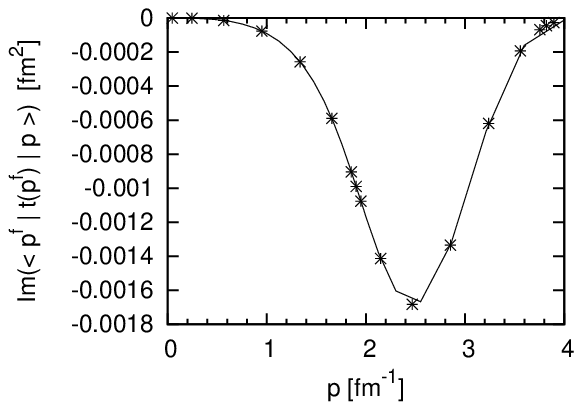}
\includegraphics[width=0.45\textwidth,angle=0]{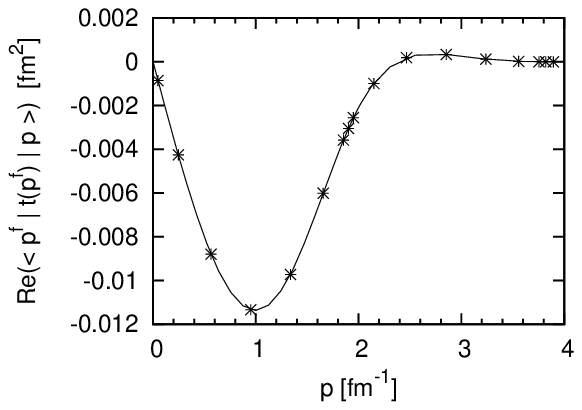}
\includegraphics[width=0.45\textwidth,angle=0]{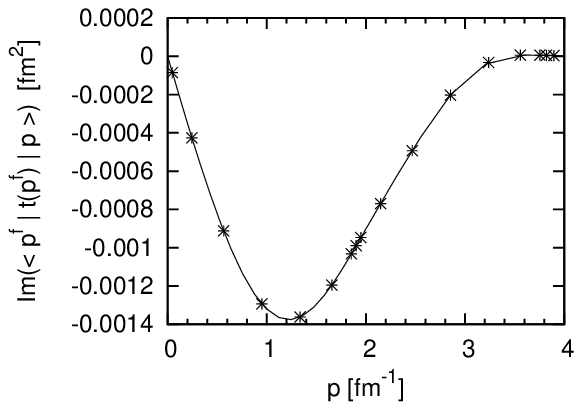}
\includegraphics[width=0.45\textwidth,angle=0]{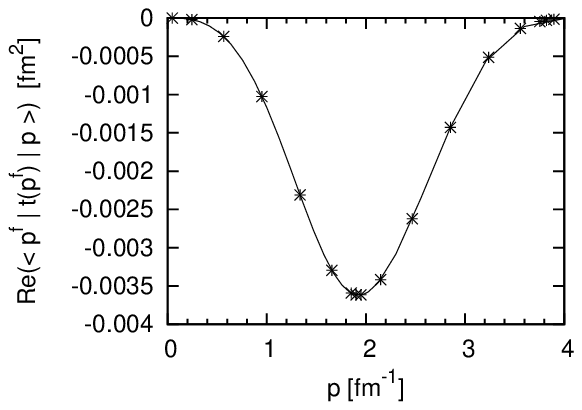}
\includegraphics[width=0.45\textwidth,angle=0]{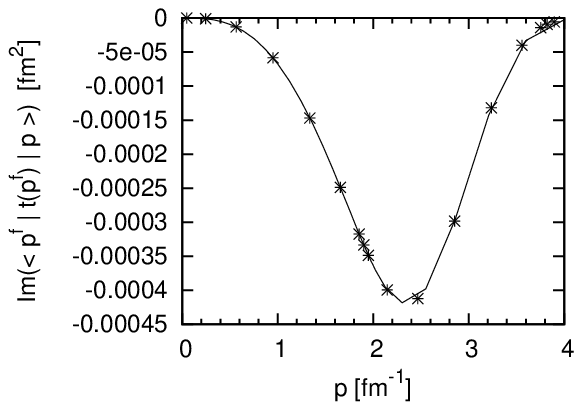}
\caption{The same as in Fig.~\ref{f3}
for the half-shell $^3P_2-^3F_2$ t-matrix. Rows
show different $l$ and $l'$ cases (from top to bottom):
($l=1$, $l'=1$), 
($l=3$, $l'=1$),
($l=1$, $l'=3$) and
($l=3$, $l'=3$).
}
\label{f5}
\end{figure}

\begin{figure}[hp]\centering
\includegraphics[width=0.45\textwidth,angle=0]{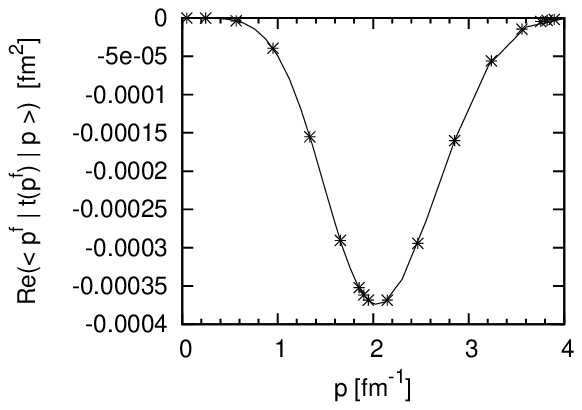}
\includegraphics[width=0.45\textwidth,angle=0]{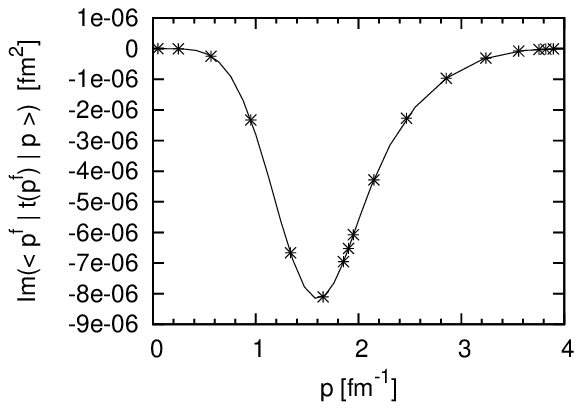}
\includegraphics[width=0.45\textwidth,angle=0]{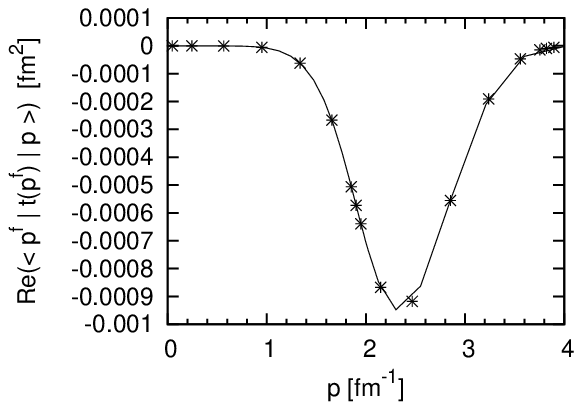}
\includegraphics[width=0.45\textwidth,angle=0]{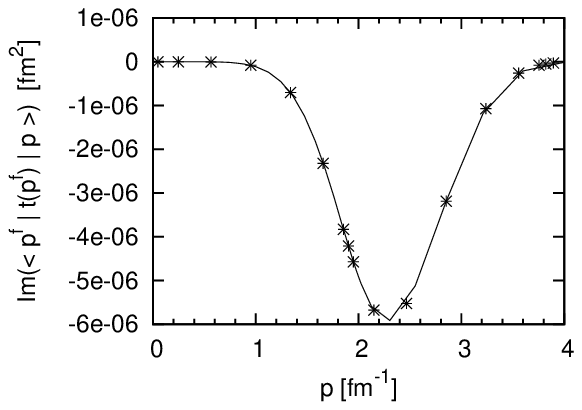}
\includegraphics[width=0.45\textwidth,angle=0]{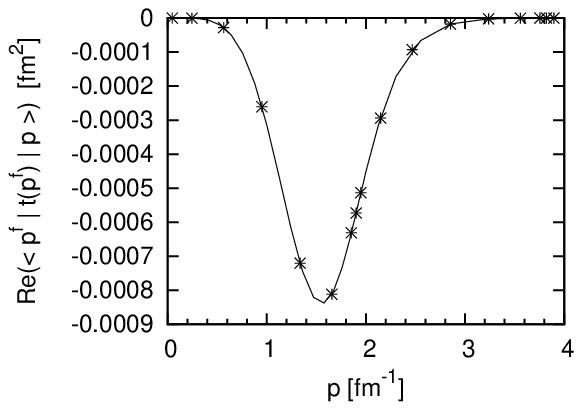}
\includegraphics[width=0.45\textwidth,angle=0]{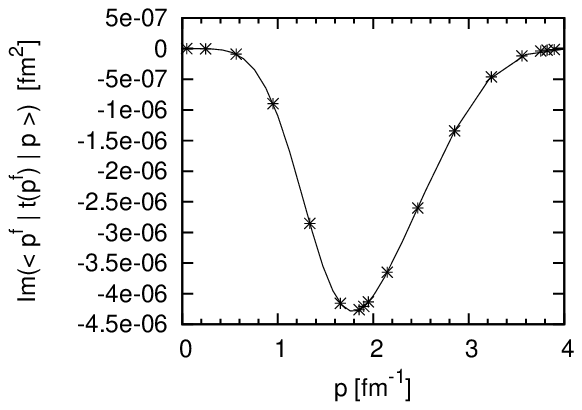}
\includegraphics[width=0.45\textwidth,angle=0]{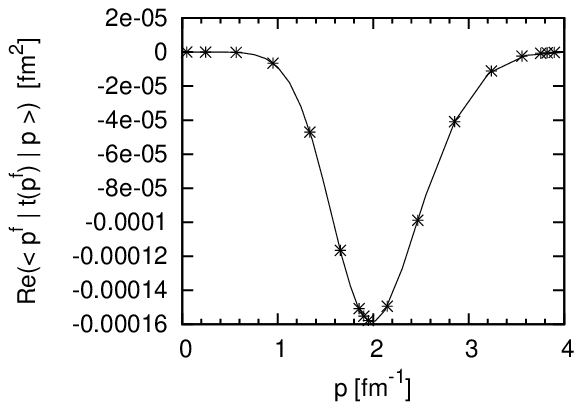}
\includegraphics[width=0.45\textwidth,angle=0]{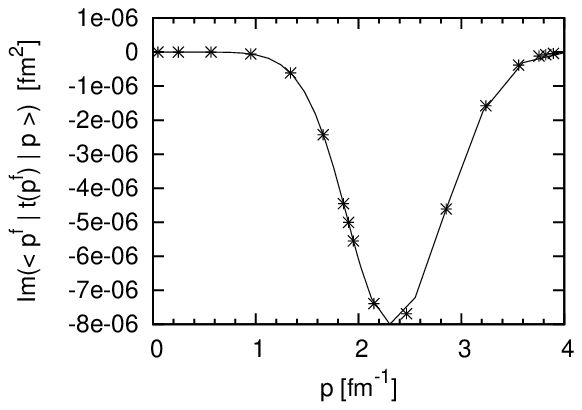}
\caption{The same as in Fig.~\ref{f3}
for the half-shell $^3H_6-^3J_6$ t-matrix. Rows
show different $l$ and $l'$ cases (from top to bottom):
($l=5$, $l'=5$), 
($l=7$, $l'=5$),
($l=5$, $l'=7$) and
($l=7$, $l'=7$).
}
\label{f6}
\end{figure}

In order to further check our t-matrix coefficients,
we used them to calculate the partial wave representation 
of the t-matrix: 
\[
\langle p^{\mathrm{f}} ( l' s ) j \mid t \left( E_{2N}= \frac{ (\mathbf{p}^{\mathrm{f}})^2 }{m}  \right) 
\mid p ( l s ) j \rangle,
\]
where $l$ ($l'$) is the initial (final) angular
momentum of the 2N system, $s$ is the 2N (conserved) spin and 
$j$ is the 2N (conserved) total angular momentum.
These matrix elements can be compared 
with direct solutions of LSE obtained in the standard partial 
wave representation. We performed the projection of the three-dimensional
t-matrix on partial waves, employing the simple method
proposed in Ref.~\cite{newpwd} for NN forces. 
In Figs.~\ref{f3}-\ref{f6} we show examples for the uncoupled and coupled 
channels, with the 2N isospin $t=0$ and $t=1$.  We chose $p^{\mathrm{f}} \approx$ 1.9 fm$^{-1}$,
which corresponds to a relatively high NN center of mass energy, $E_{2N}$= 150 MeV.
For such an energy many partial waves contribute to the NN scattering observables
and the question arises if the partial contributions are consistent
with the full three-dimensional calculations. From Figs.~\ref{f3}-\ref{f6} 
(and many other cases which are not shown here) we infer that this is really
the case. The agreement between results based on the two quite different
approaches is very good.

The final ingredient in our framework is the 2N current operator. It consists
of the single-nucleon and two-nucleon operators. For the purpose 
of this paper we assume that its single-nucleon part comprises the standard nonrelativistic 
charge density as well as the convection and spin current operators. In the 2N part we take
for simplicity only the leading one-pion-exchange current operator in the chiral 
effective field theory representation. In our three-dimensional treatment 
of the $ ^2{\rm H} ( e, e' p) n $ reaction, we calculate the spin and isospin 
matrix elements of the current operator directly, using simple matrix representations 
of the spin and isospin operators and the concept of the Kronecker product to deal
with the two-nucleon spin and isospin spaces. In the traditional calculations,
a partial wave decomposition of the current operator is required. 
It is a rather easy task for the single-nucleon part of the current.
For the one-pion-exchange current operator it is known analytically 
(see for instance Ref.~\cite{kotlyar}). It can also be obtained using the method 
proposed in Refs.~\cite{rozpedzik1,rozpedzik2}, where even more complicated 
two-pion-exchange current operators were considered.

To give examples of our results on deuteron electro-disintegration we chose several electron kinematics 
given in Table~\ref{6kinematics}. They allowed us to study the reaction
for three different internal nucleon-nucleon energies (corresponding to the 
three values of $p^{\mathrm{f}}$) and for five values of the three-momentum transfer 
$Q$. The first parameter is the input for the t-matrix calculations 
and the second one specifies the properties of the current matrix elements.

\begin{table}[hp]\centering
\caption{\label{6kinematics} The six electron kinematics 
considered in the paper for the exclusive 
$ ^2{\rm H} ( e, e' p) n $ process. 
The initial electron energy ($E_{e}$), the electron
scattering angle ($\theta_e$), the final electron energy ($E_{e}'$),
the final relative nucleon-nucleon momentum ($|\mathbf{p^{\mathrm{f}}}|$), the energy transfer
($\omega$), and the magnitude of the three-momentum transfer ($Q$) are given. 
}
\begin{tabular}{|c|c|c|c|c|c|c|}
\hline
 &  $E_{e}$  & $\theta_e$  & $E_{e}'$ &  $p^{\mathrm{f}}$ of Eq. (\ref{relmom})  & $\omega = E_{e} - E_{e}'$ & $Q$ of Eq. (\ref{momentumQ}) \cr
 &  MeV & deg         &  MeV &  MeV/c &  MeV     &  MeV/c \cr
 \hline
$K1$ &  500 &  6.9 &490.3  & 78.1 &  9.7&  60 \cr
$K2$ &  500 & 17.4 &485.3  & 78.1 & 14.7& 150 \cr
$K3$ &  500 &  6.1 &467.0 & 158.7 & 30.0 & 60 \cr
$K4$ &  500 & 36.4 &447.0 & 158.7 & 53.0 &300 \cr
$K5$ &  500 & 16.3 &337.1 & 375.3 &162.9 &200 \cr
$K6$ &  500 & 73.7 &281.2 & 375.3 &218.8 &500 \cr
\hline
\end{tabular}
\end{table}

\begin{figure}[hp]\centering
\includegraphics[width=4cm,angle=0]{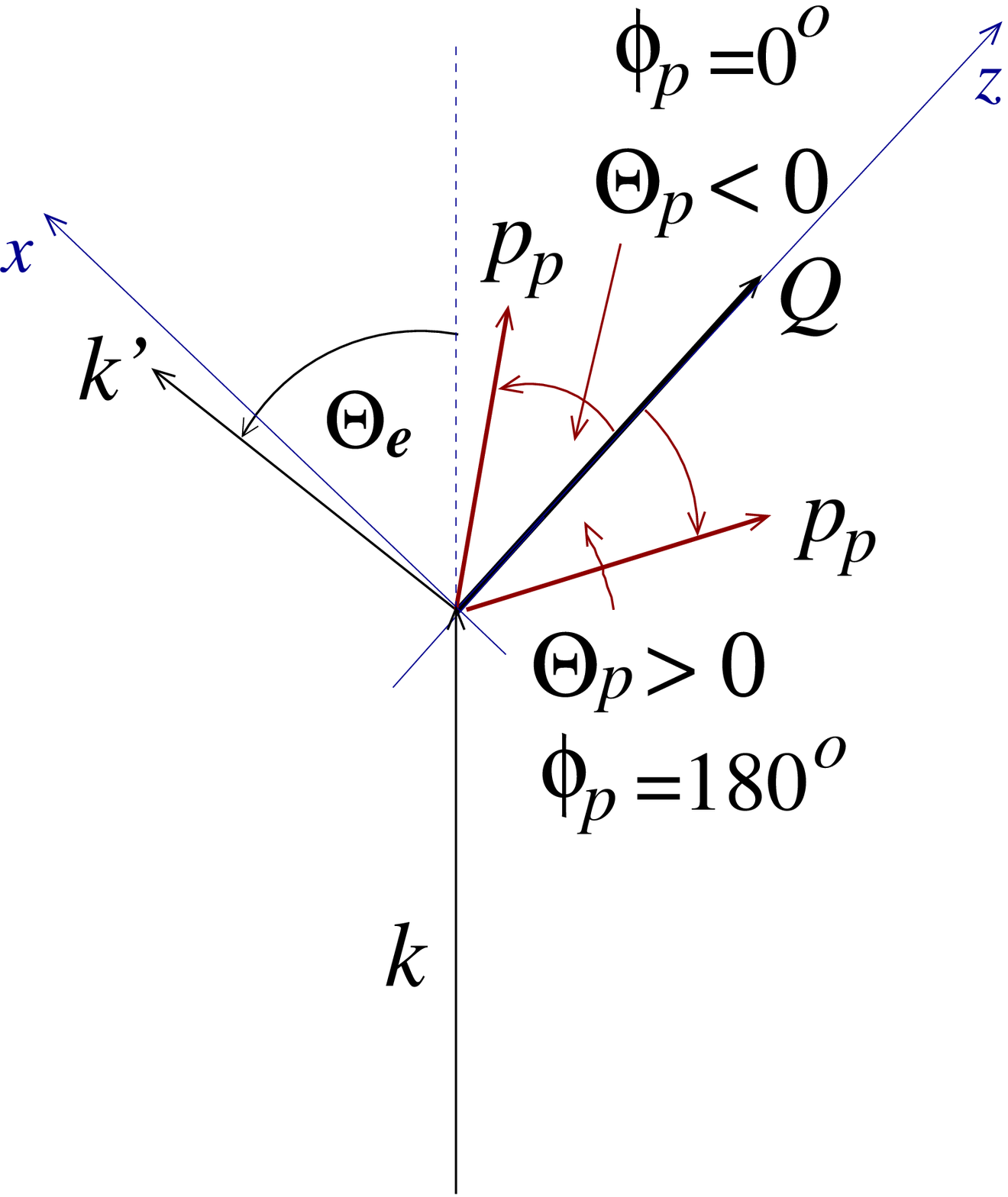}
\caption{The kinematics for the exclusive $ ^2{\rm H} ( e, e' p) n $ process.
$\vec k$ ($\vec k'$) is the initial (final) electron momentum.
We neglect the electron mass, so $\mid \vec k \mid = E $ and  $\mid \vec k' \mid = E' $.}
\label{f7}
\end{figure}

In addition to the information given in Table~\ref{6kinematics},
we need to label the exclusive kinematics. For the fixed "electron arm",
we deal in fact with a two-body kinematics in the final proton-neutron
system. We restrict ourselves here to the case where the proton is ejected in the electron plane, 
where $\theta_p$ would be the angle between the three-momentum transfer $\vec Q$ 
and the final proton momentum $\vec p_p$.
Since we have to distinguish between the $\phi_p$=0 deg and $\phi_p$=180 deg
cases, we ascribe the negative sign to $\theta_p$ for $\phi_p$=0 deg.
This is shown in Fig.~\ref{f7}. 
Note that the six electron kinematics
provide a unique solution for any $\theta_p$ value and that 
$\theta_p$ changes from 0 do 180 degrees. 

We are now ready to show our results for several selected observables. 
We chose first of all the unpolarized cross section,
$d^5\sigma/( dE_{e'}\; d\Omega_{e'}\; d\Omega_{p}\; )$.
We take also into account 
one example of the spin-dependent helicity asymmetry,
\[
 A_\parallel \equiv \frac{ \sigma (h=+1, {\vec J}_d ) - \sigma (h=-1, {\vec J}_d ) }
                         { \sigma (h=+1, {\vec J}_d ) + \sigma (h=-1, {\vec J}_d ) } \,  
\]
where $h$ is the initial electron helicity and the projection of the initial deuteron total angular momentum ($\mathbf{J}_{d}$) on $\vec Q$, $ J_{dz}$, is equal 1.
In addition we show our predictions for the deuteron tensor analyzing powers $T_{kq}$. Note that 
they are calculated in the system, where $\vec Q \parallel \hat z$. 

\begin{figure}[hp]\centering
\includegraphics[width=0.45\textwidth,angle=0]{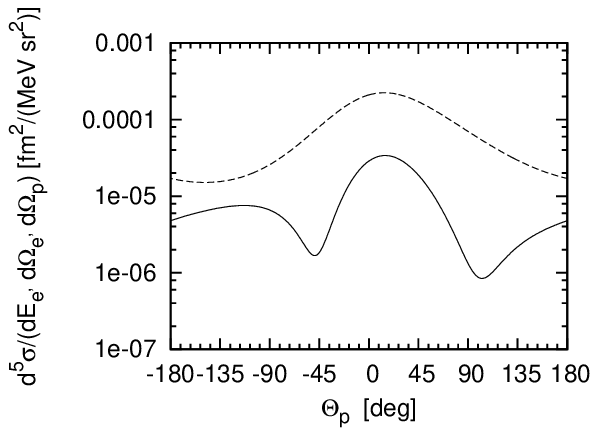}
\includegraphics[width=0.45\textwidth,angle=0]{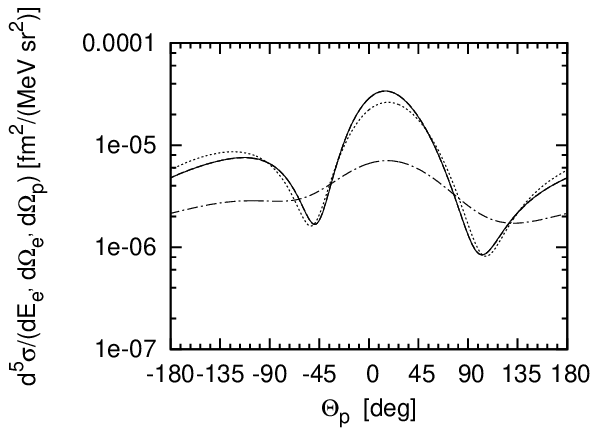}
\caption{The unpolarized cross section $d^5\sigma/( dE_{e'}\; d\Omega_{e'}\; d\Omega_{p}\; )$
as a function of the proton scattering angle $\theta_p$
for the $K3$ electron kinematics from Table~\ref{6kinematics}.
In the left panel plane wave results (dashed line) are compared
with results of the full calculations (solid line) obtained within
the same three-dimensional scheme.
In the right panel convergence of the full results calculated
with a different number of nucleon-nucleon partial waves
towards the full three-dimensional
prediction (solid line) is shown.
Partial wave based results with
$ j \le 1 $ (dash-dotted line),
$ j \le 2 $ (dotted line) and
$ j \le 4 $ (dashed line) are displayed.}
\label{f14}
\end{figure}

\begin{figure}[hp]\centering
\includegraphics[width=0.45\textwidth,angle=0]{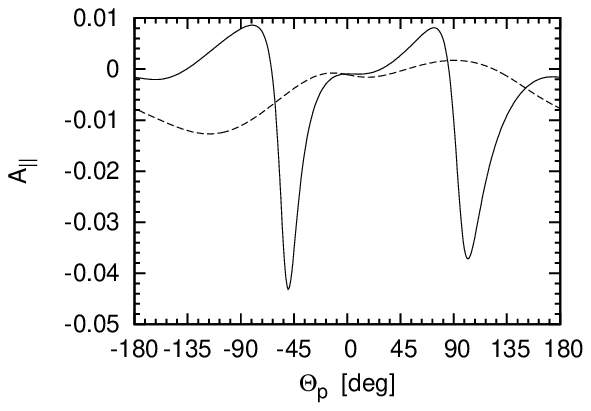}
\includegraphics[width=0.45\textwidth,angle=0]{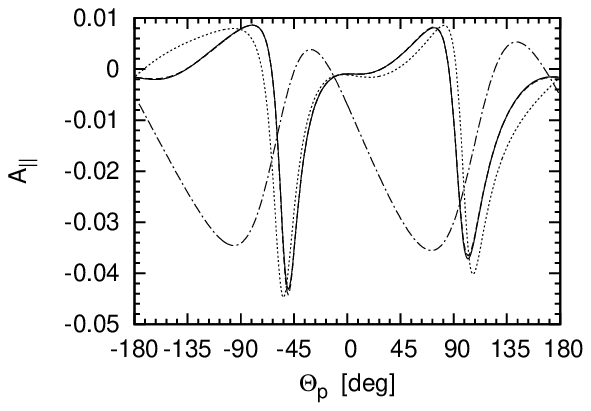}
\caption{The same as in Fig.~\ref{f14} for the spin-dependent
helicity asymmetry $A_\parallel$.}
\label{f15}
\end{figure}

\begin{figure}[hp]\centering
\includegraphics[width=0.45\textwidth,angle=0]{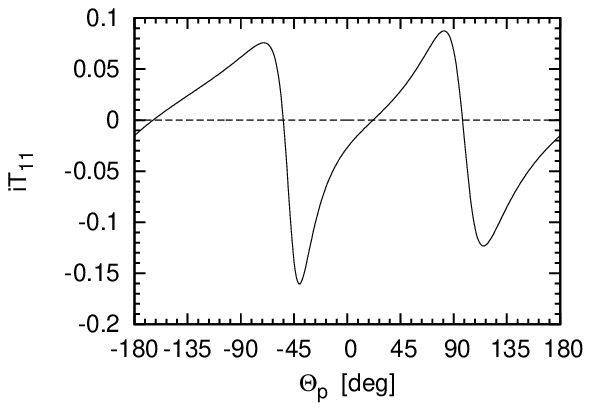}
\includegraphics[width=0.45\textwidth,angle=0]{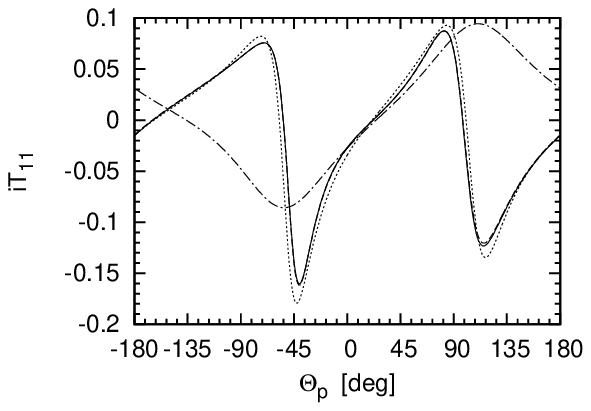}
\includegraphics[width=0.45\textwidth,angle=0]{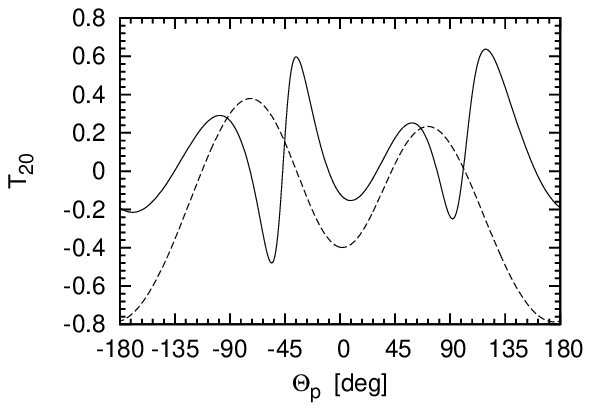}
\includegraphics[width=0.45\textwidth,angle=0]{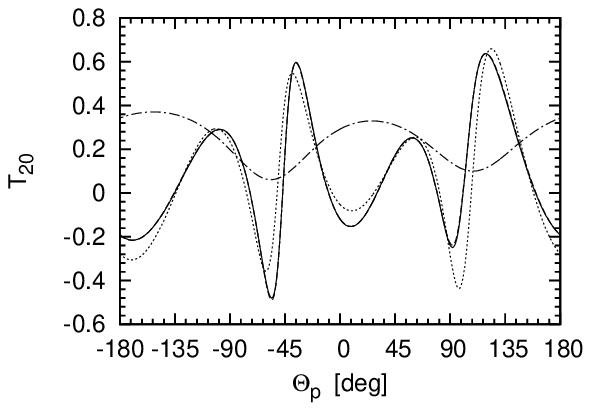}
\includegraphics[width=0.45\textwidth,angle=0]{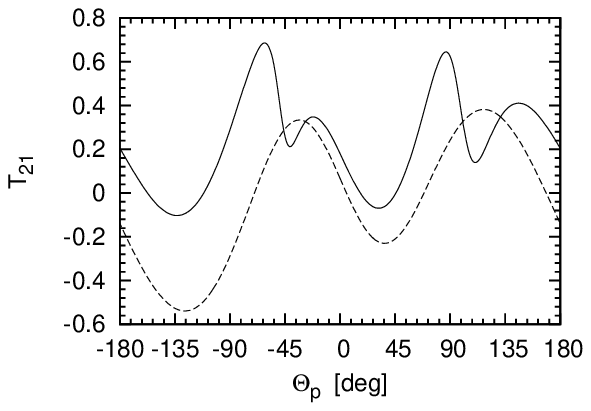}
\includegraphics[width=0.45\textwidth,angle=0]{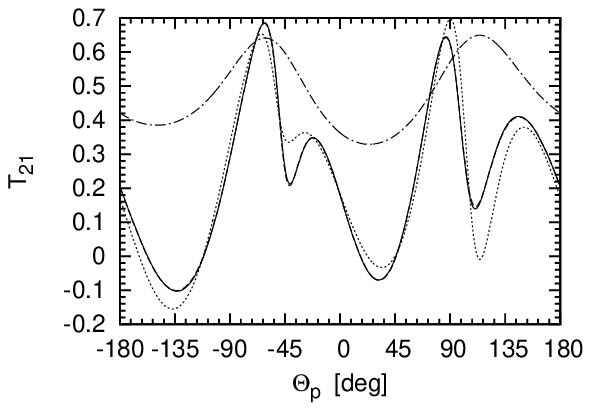}
\includegraphics[width=0.45\textwidth,angle=0]{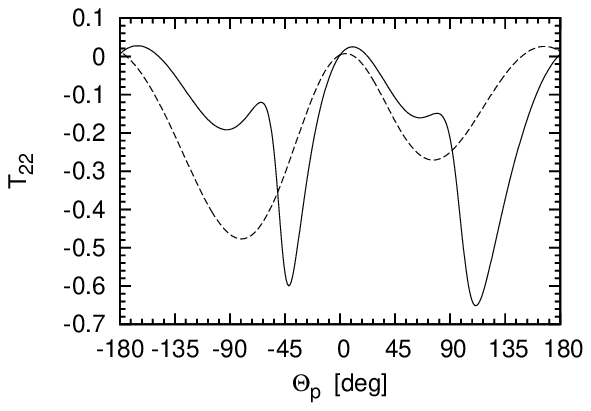}
\includegraphics[width=0.45\textwidth,angle=0]{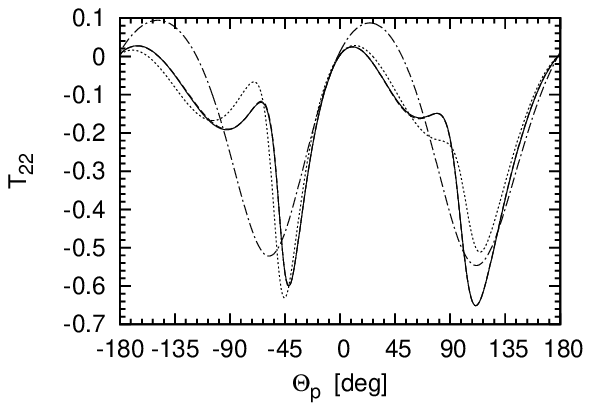}
\caption{The same as in Fig.~\ref{f14} for the
deuteron analyzing powers $T_{kq}$.}
\label{f16}
\end{figure}

\begin{figure}[hp]\centering
\includegraphics[width=0.45\textwidth,angle=0]{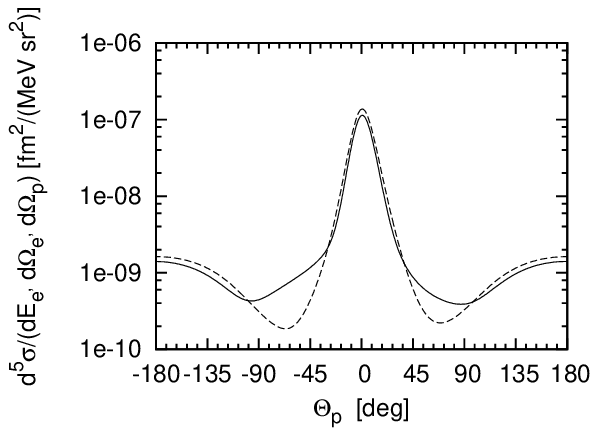}
\includegraphics[width=0.45\textwidth,angle=0]{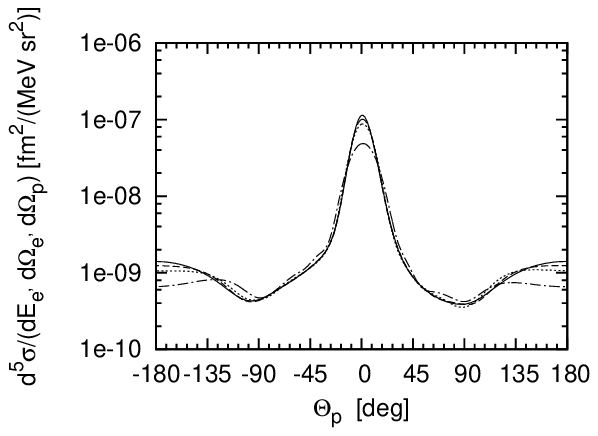}
\caption{The same as in Fig.~\ref{f14} for the $K6$ electron 
kinematics from Table~\ref{6kinematics}.
In the right panel
partial wave based results with
$ j \le 4 $ (dash-dotted line),
$ j \le 7 $ (dotted line) and
$ j \le 9 $ (dashed line) are compared
with the full three-dimensional
prediction (solid line).}
\label{f23}
\end{figure}

\begin{figure}[hp]\centering
\includegraphics[width=0.45\textwidth,angle=0]{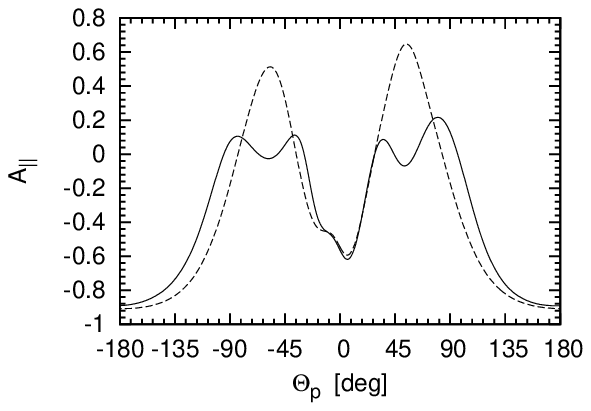}
\includegraphics[width=0.45\textwidth,angle=0]{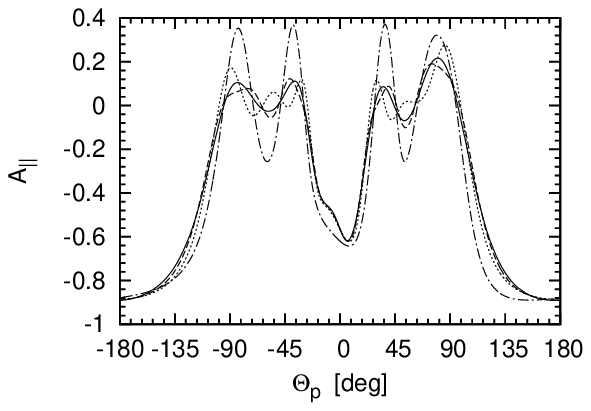}
\caption{The same as in Fig.~\ref{f23} for the spin-dependent
helicity asymmetry $A_\parallel$.}
\label{f24}
\end{figure}

\begin{figure}[hp]\centering
\includegraphics[width=0.45\textwidth,angle=0]{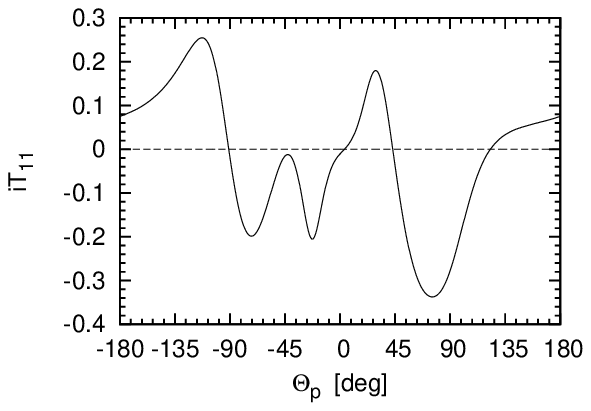}
\includegraphics[width=0.45\textwidth,angle=0]{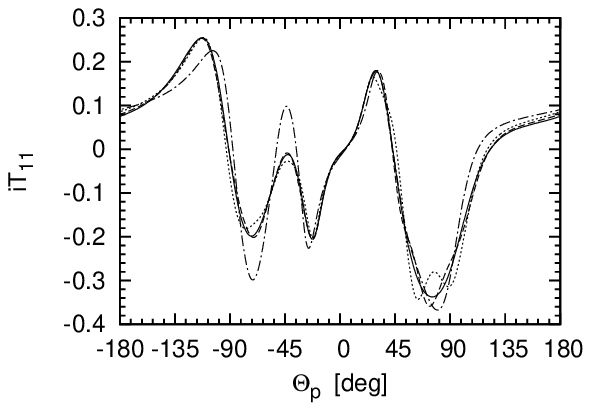}
\includegraphics[width=0.45\textwidth,angle=0]{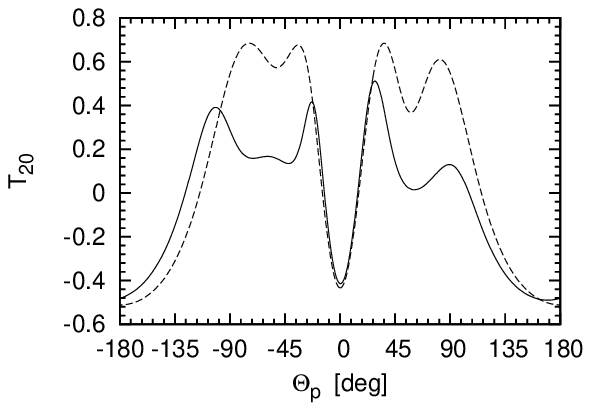}
\includegraphics[width=0.45\textwidth,angle=0]{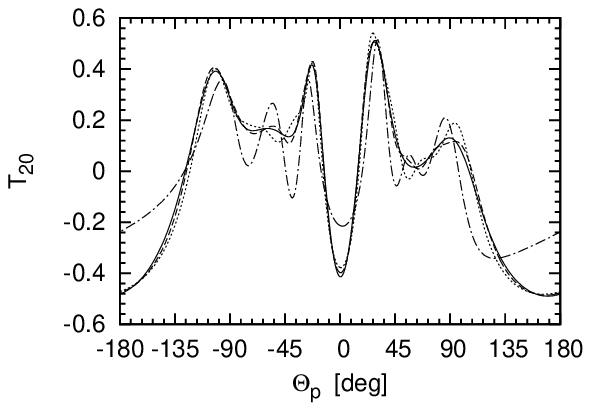}
\includegraphics[width=0.45\textwidth,angle=0]{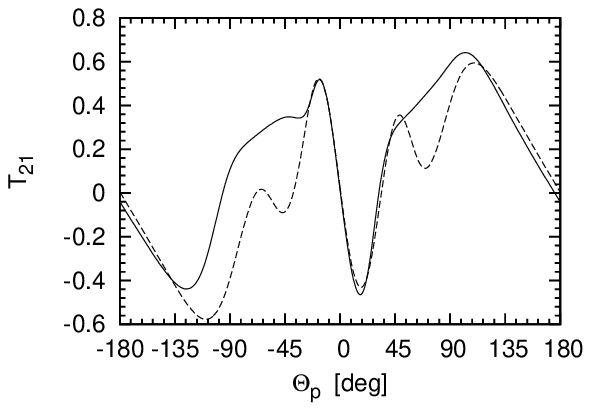}
\includegraphics[width=0.45\textwidth,angle=0]{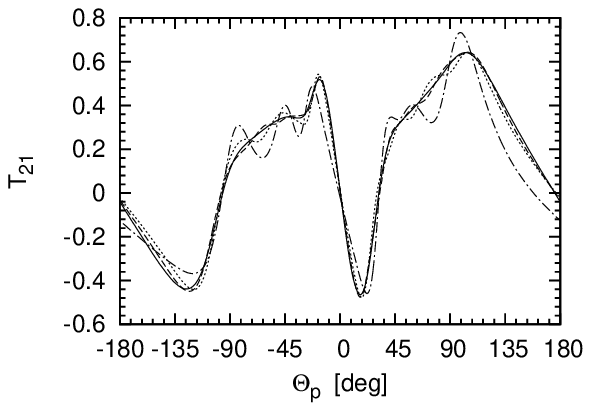}
\includegraphics[width=0.45\textwidth,angle=0]{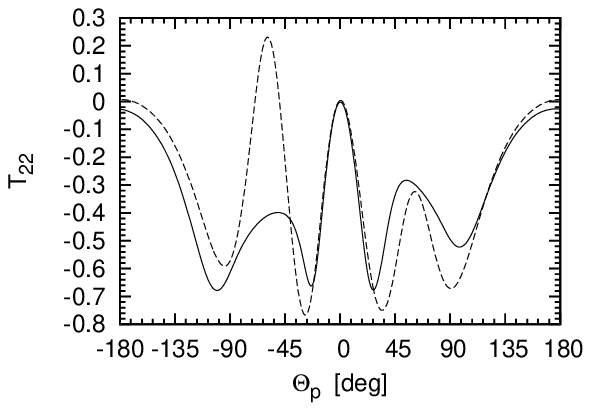}
\includegraphics[width=0.45\textwidth,angle=0]{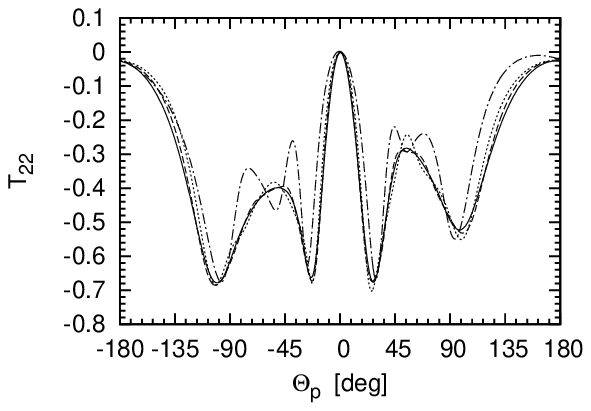}
\caption{The same as in Fig.~\ref{f23} for the
deuteron analyzing powers $T_{kq}$.}
\label{f25}
\end{figure}

Our primary goal was to compare results based on the partial wave decomposition for 
the t-matrix and the nuclear current operator with new predictions resulting from the 
three-dimensional scheme. We observed 
a perfect agreement for all the electron kinematics and for all
the considered observables, if a sufficient number
of partial waves in the first type of calculations is included.
The six kinematics can be divided into two groups: ($K1$ , $K3$ , $K5$) and ($K2$ , $K4$ , $K6$). In each group a similar type of convergence of the observables with respect to the number of partial waves is observed.
That is why 
in Figs.~\ref{f14}--\ref{f25} we show predictions for two representative 
($K3$ and $K6$) kinematics only. In the first case we see a rapid convergence 
and partial wave based results with $ j \le 4$ are already very close 
to the full three-dimensional prediction. In the second case all partial waves
with $ j \le 9$ are necessary to achieve convergence. 

\begin{figure}[hp]\centering
\includegraphics[width=0.45\textwidth,angle=0]{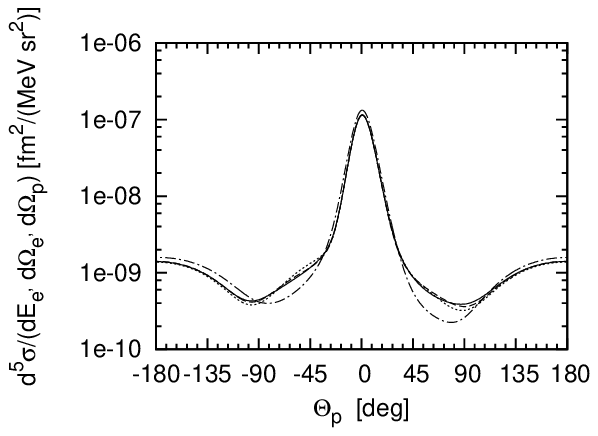}
\includegraphics[width=0.45\textwidth,angle=0]{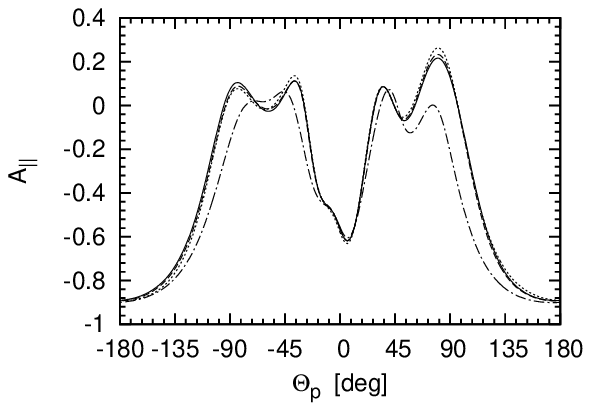}
\includegraphics[width=0.45\textwidth,angle=0]{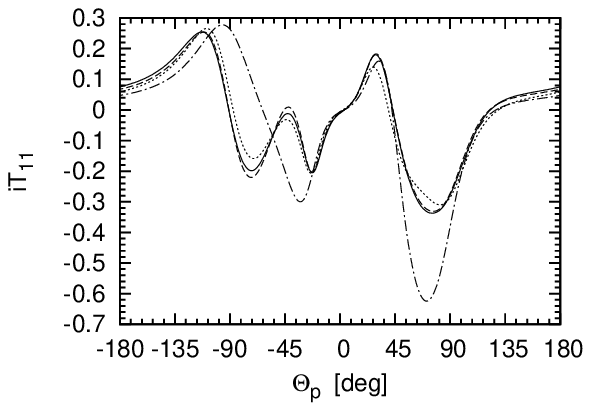}
\includegraphics[width=0.45\textwidth,angle=0]{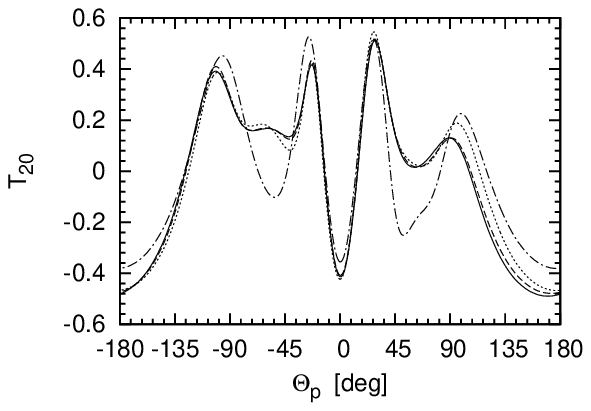}
\includegraphics[width=0.45\textwidth,angle=0]{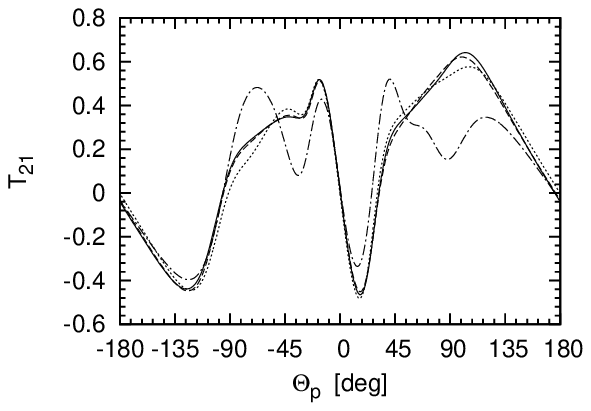}
\includegraphics[width=0.45\textwidth,angle=0]{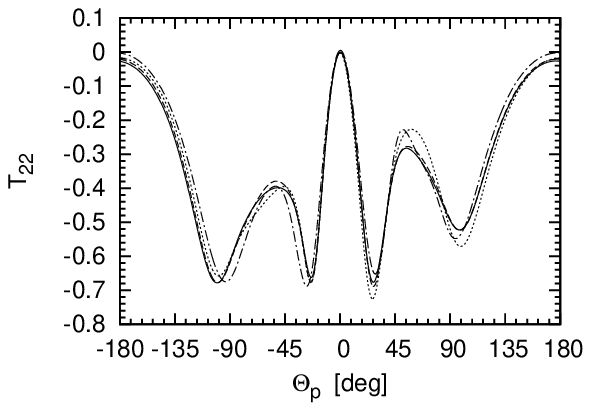}
\caption{Various observables for the $K6$ electron 
kinematics from Table~\ref{6kinematics}. The single-nucleon current contribution
to the plane wave part of the nuclear matrix element is now calculated 
without partial wave decomposition. The remaining parts of the nuclear 
matrix element are still calculated using
partial waves. Results obtained with  
$ j \le 1 $ (dash-dotted line),
$ j \le 2 $ (dotted line) and
$ j \le 3 $ (dashed line) are compared
with the full three-dimensional
prediction (solid line).}
\label{f26}
\end{figure}

It is interesting to see that slow convergence for the
$K2$, $K4$ and $K6$ kinematics does not result from the higher 
$p^{\mathrm{f}}$ values (that is from the t-matrix)
but is related to the $Q$ values and thus to the partial wave decomposition 
of the current operator. It is well known (see for example Ref.~\cite{rozpedzik1}) 
that especially the partial wave decomposition of the single nucleon current 
requires many partial waves. However, even if the initial bound state 
is given in the partial wave representation, the single nucleon 
current can be applied directly in the case of the plane wave amplitudes.
This holds not only for the two- but also for the three-nucleon 
system~\cite{report2005}. In order to demonstrate this behaviour, 
we showed in Fig.~\ref{f26} observables for the $K6$ kinematics.
In this case the single-nucleon current contribution to
the plane wave amplitude is calculated without partial wave 
decomposition. We clearly see that the convergence is significantly 
improved, even if the two-nucleon current
contribution to the plane wave part of the nuclear matrix element
and the whole rescattering part of the nuclear matrix element
is calculated with the partial wave decomposition.

\clearpage

\clearpage

\clearpage

\section{Conclusions and outlook}

The presented method to treat several electroweak processes involving 1N and 2N current operators in three dimensions can successfully replace standard partial wave treatment. We showed, for the case of electron induced deuteron disintegration, that results obtained using the new approach agree very well with those obtained using PWD. For all observables considered in this paper, the traditional results converge to the three dimensional predictions when the number of partial waves is sufficiently high (Figs.~\ref{f14}--\ref{f25}). 

Our formalism employs the two nucleon bound state,
the 2N $t$ matrix
and the current operators in the joined isospin - spin space of the 2N system using the three dimensional formalism in 2N momentum space. Each element of this framework has been separately tested and compared with the standard PWD approach. Our method seems to be more flexible and can deal with the rich structures of the 2N force and the current operator, especially derived within the higher orders of the chiral effective field theory \cite{Epelbaum:2004fk,2Ncurrents,2Ncurrents1}. We plan to use our framework for other processes such as muon capture or neutrino induced deuteron disintegration. In \cite{Marcucci:2010ts} muon capture on $^{2}$H and $^{3}$He was considered using the PWD approach. It would be interesting to compare those results with three dimensional calculations. A similar convergence to three dimensional results as in Figs.~\ref{f14}--\ref{f25} is expected.

\section*{Acknowledgements}

We acknowledge support by the Foundation for Polish Science - MPD program, co-financed by the European Union within the Regional Development Fund. This work was supported by the Polish National Science Center under Grant No. DEC-2011/01/B/ST2/00578 and partially by the EU HadronPhysics3 project "Exciting Physics Of Strong Interactions".

One of the authors (JG) would like to thank K. Sagara 
for the hospitality extended to him during his stay
at the Kyushu University and E. Epelbaum 
for the hospitality extended to him during his stay
at the Ruhr-Universit\"at, Bochum.
KT would like to thank Ulf-G. Mei{\ss}ner 
for the hospitality extended to him during his stay at the Institut f\"{u}r Kernphysik in the Forschungszentrum J\"{u}lich.
The numerical calculations have been partly performed on the supercomputers of the JSC, J{\"u}lich, Germany.

\bibliographystyle{unsrt}

\bibliography{bibl}

\end{document}